\documentclass[apj, twocolappendix, numberedappendix]{emulateapj}

\usepackage[colorlinks, linkcolor=red, anchorcolor=blue, citecolor=green]{hyperref}
\usepackage{makeidx}

\usepackage{natbib}

\usepackage{graphicx}

\begin{document}

\title{Can We Detect the Color--Density Relation with Photometric Redshifts?}

\author{Chuan-Chin Lai\altaffilmark{1, 2}, Lihwai Lin\altaffilmark{2}, Hung-Yu Jian\altaffilmark{2, 3}, Tzi-Hong Chiueh\altaffilmark{1, 3}, Alex Merson\altaffilmark{4}, Carlton M. Baugh\altaffilmark{5}, Sebastien Foucaud \altaffilmark{6}, Chin-Wei Chen \altaffilmark{2}, Wen-Ping Chen \altaffilmark{7}}

\email{chuanchinlai@gmail.com}
\altaffiltext{1}{Graduate Institute of Astrophysics, National Taiwan University, Taipei, 10617, Taiwan (R.O.C.)}
\altaffiltext{2}{Institute of Astronomy \& Astrophysics, Academia Sinica, Taipei 10617, Taiwan (R.O.C.)}
\altaffiltext{3}{Department of Physics, National Taiwan University, Taipei, 10617, Taiwan (R.O.C.)}
\altaffiltext{4}{Department of Physics and Astronomy, University College London, Gower Street, London WC1E 6BT, UK}
\altaffiltext{5}{Institute for Computational Cosmology, Department of Physics, Durham University, South Road, Durham DH1 3LE, UK}
\altaffiltext{6}{Center for Astronomy and Astrophysics, Shanghai Jiao Tong University, Shanghai 200240, China}
\altaffiltext{7}{Graduate Institute of Astronomy, National Central University, Chung-Li 32054, Taiwan (R.O.C.)}

\begin{abstract}

A variety of methods have been proposed to define and to quantify galaxy environments. While these techniques work well in general with spectroscopic redshift samples, their application to photometric redshift surveys remains uncertain. To investigate whether galaxy environments can be robustly measured with photo-z samples, we quantify how the density measured with the nearest neighbor approach is affected by photo-z uncertainties by using the Durham mock galaxy catalogs in which the 3D real-space environments and the properties of galaxies are exactly known. Furthermore, we present an optimization scheme in the choice of parameters used in the 2D projected measurements which yield the tightest correlation with respect to the 3D real-space environments. By adopting the optimized parameters in the density measurements, we show that the correlation between the 2D projected optimized density and real-space density can still be revealed, and the color--density relation is also visible out to $z \sim 0.8$ even for a photo-z uncertainty ($\sigma_{\Delta_{z}/(1+z)}$) up to 0.06. We find that at the redshift $0.3 < z < 0.5$ a deep ($i \sim 25$) photometric redshift survey with $\sigma_{\Delta_{z}/(1+z)} = 0.02$ yields a comparable performance of small-scale density measurement to a shallower $i \sim$ 22.5 spectroscopic sample with $\sim$ 10\% sampling rate. Finally, we discuss the application of the local density measurements to the Pan-STARRS1 Medium Deep survey, one of the largest deep optical imaging surveys. Using data from $\sim5$ square degrees of survey area, our results show that it is possible to measure local density and to probe the color--density relation with 3$\sigma$ confidence level out to $z \sim 0.8$ in the PS-MDS. The color--density relation, however, quickly degrades for data covering smaller areas.

\end{abstract}

\keywords{galaxies: evolution -- galaxies: formation -- galaxies: halos}

\section{Introduction}

\label{sec:1}

Recent observations have shown that various galaxy properties such as star formation rate, color and morphology are strongly correlated with galaxy environment \citep{2004ApJ...601L..29H, 2006ApJ...645..977B, 2006ApJ...647L..21H, 2006MNRAS.370..198C, 2007MNRAS.376.1445C, 2007ApJS..172..284C, 2007A&amp;A...468...33E, 2009ApJ...690.1883G, 2012MNRAS.419.3018C, 2013ApJ...767...89M, 2014ApJ...782...33L, 2014ApJ...796...51D}. These studies indicate that galaxies located in dense environments, such as galaxy groups and clusters, tend to be redder, elliptical and with lower star formation rates. Several physical processes, including ram pressure stripping \citep{1972ApJ...176....1G, 2000Sci...288.1617Q}, high speed galaxy encounters \citep[galaxy harassment; ][]{1996Natur.379..613M}, galaxy-galaxy mergers \citep{1994ApJ...431L...9M}, and removal of warm and hot gas \citep[strangulation; ][]{1980ApJ...237..692L, 2000ApJ...540..113B, 2008MNRAS.383..593M} have been proposed to explain the observed relation between environment and galaxy properties. Yet exactly how the environment affects the evolution of galaxies and how important it is as opposed to internal properties of galaxies (e.g., stellar mass) is still unclear. Part of the discrepancy between previous studies may come from the differences in the sample selection as well as the definition of environment, which make the comparisons non-trivial. 
 
One of the common approaches to characterize the galaxy environment is to use the local overdensity of matter. For the rest of this paper, we refer the galaxy environment to the observed overdensity of galaxies as a proxy for this. A variety of methods have been used to define galaxy density field, for example, (1) the Fixed Aperture method which counts the number of neighbor galaxies in a fixed volume around each galaxy \citep[e.g.,][]{2009ApJ...690.1883G, 2011MNRAS.411..929G}; (2) the Annulus method which counts the number of neighbor galaxies within a circular ring around each galaxy \citep[e.g.,][]{2010MNRAS.406.1701W}; and (3) the $N^{\rm th}$ nearest neighbor that defines the local density by finding the distance from the individual reference galaxies to the $N^{\rm th}$ nearest galaxy \citep[e.g.,][]{1985ApJ...298...80C, 2003ApJ...584..210G, 2006MNRAS.373..469B, 2012MNRAS.419.2133H}. A fundamental and crucial quantity for these methods to work is the redshift information, which provides the information about the line-of-sight separation (in the absence of peculiar velocities) of two given galaxies. Observationally, there are two types of redshift that are used widely: spectroscopic redshifts and photometric redshifts (hereafter spectral-z and photo-z respectively). While the spectral-z samples have greater precision in the redshift measurement, they suffer from incompleteness and are observationally expensive for high-redshift galaxies. 

\begin{figure}
\epsscale{1.0}
\plotone{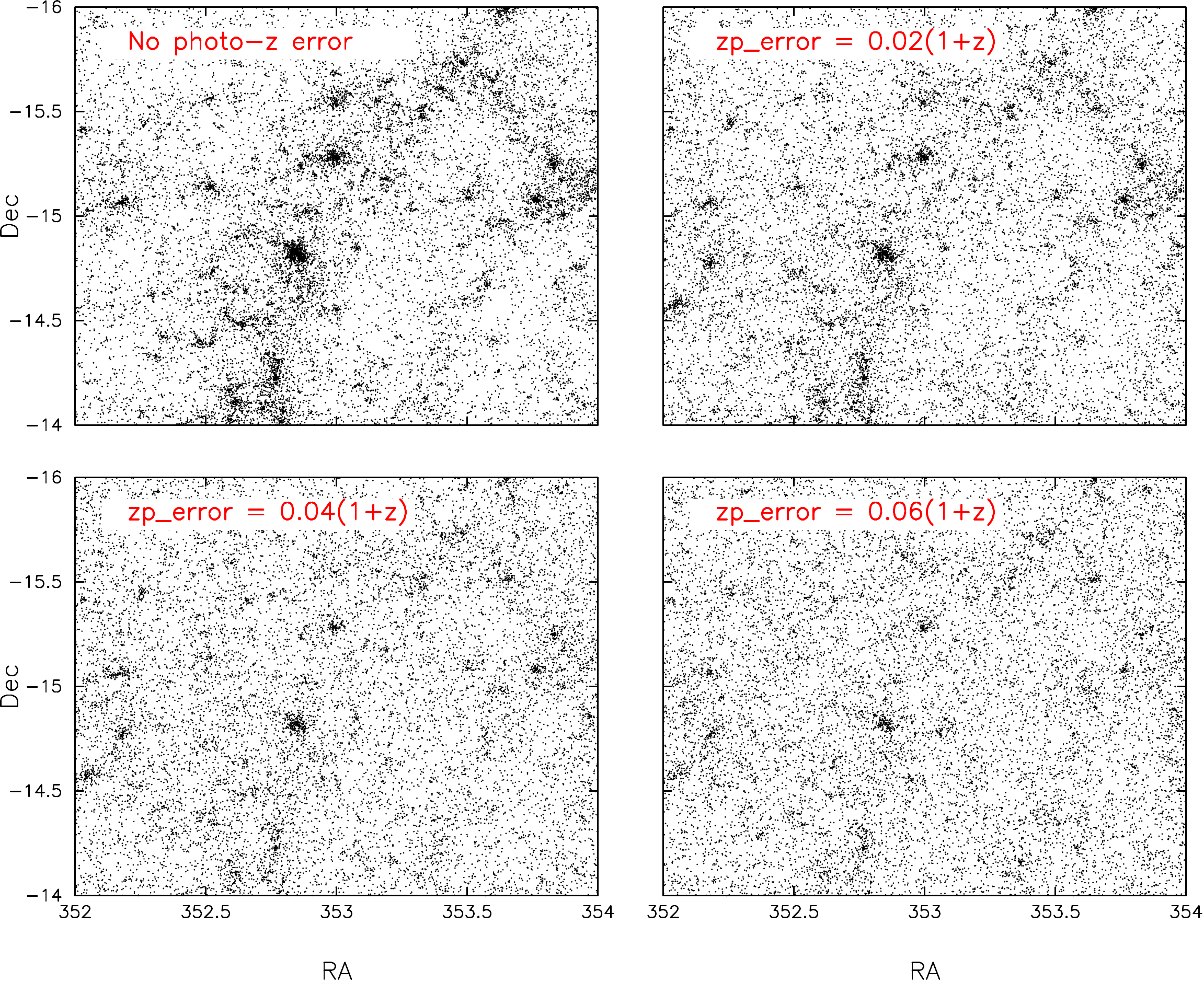}
\caption{Spatial distribution of mock galaxies projected onto the plane of the sky with redshifts perturbed corresponding to different photo-z errors, 0.00, $0.02(1+z)$, $0.04(1+z)$ and $0.06(1+z)$ at $0.3 < z_{\rm photo} < 0.35$. \label{fig:1} }
\end{figure}

\begin{figure}
\epsscale{1.0}
\plotone{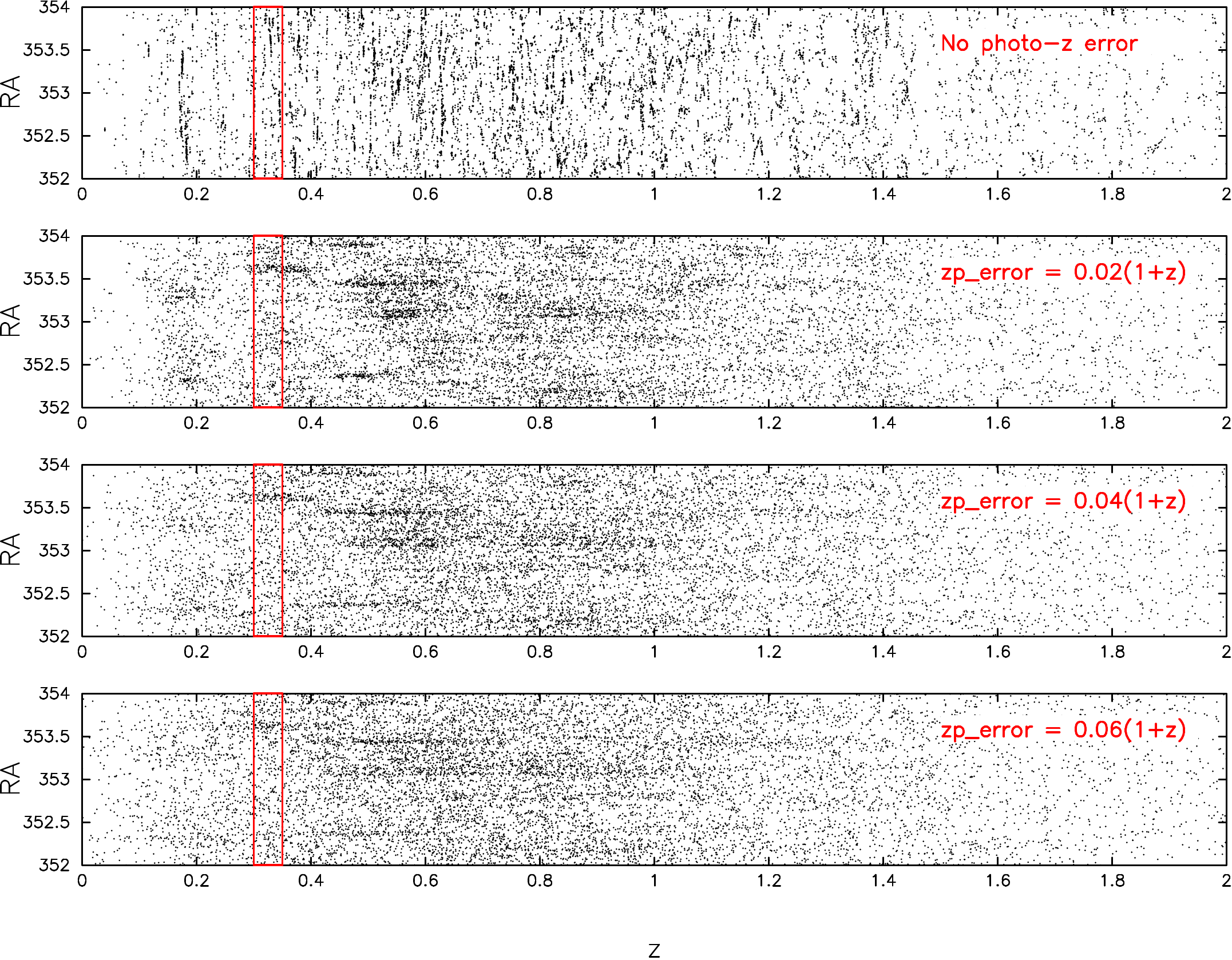}
\caption{Spatial distribution of mock galaxies seen in one line-of-sight projection and redshift. A 0.05 degrees interval in DEC is used when projecting galaxies onto the plane. The redshift of galaxies are perturbed according to different photo-z errors, 0.00, $0.02(1+z)$, $0.04(1+z)$ and $0.06(1+z)$. The red rectangle indicates the redshift range $0.3 < z_{\rm photo} < 0.35$ used in Figure \ref{fig:1}. \label{fig:21} }
\end{figure}

\begin{figure}
\epsscale{1.0}
\plotone{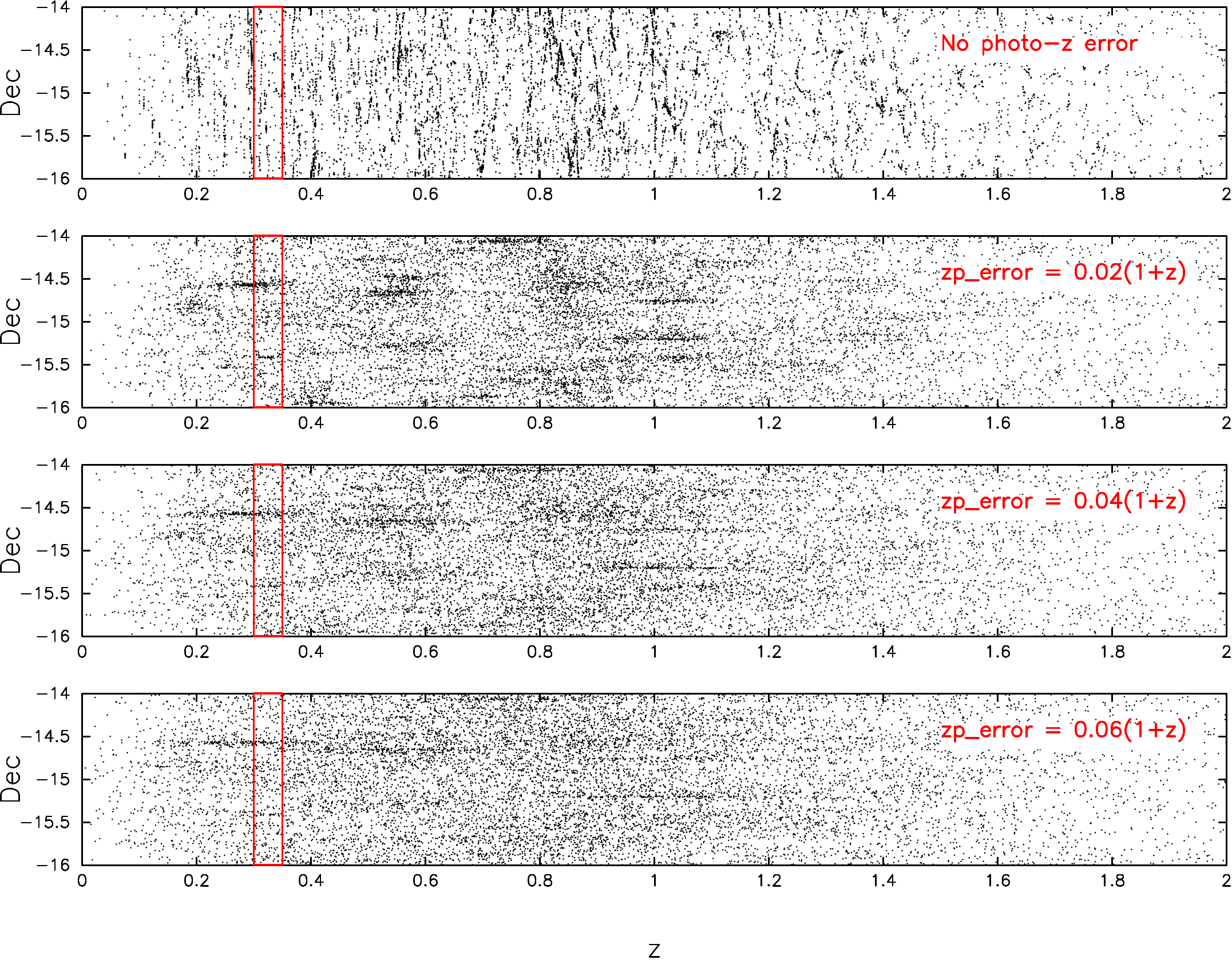}
\caption{Spatial distribution of mock galaxies seen in one line-of-sight projection and redshift. A 0.05 degrees interval in RA is used when projecting galaxies onto the plane. The redshift of galaxies are perturbed according to different photo-z errors, 0.00, $0.02(1+z)$, $0.04(1+z)$ and $0.06(1+z)$. The red rectangle indicates the redshift range $0.3 < z_{\rm photo} < 0.35$ used in Figure \ref{fig:1}. \label{fig:22} }
\end{figure}

To date, environment studies using large spectroscopic surveys such as SDSS \citep{2000AJ....120.1579Y}, DEEP2 \citep{2003SPIE.4834..161D, 2013ApJS..208....5N} and zCOSMOS \citep{2007ApJS..172...70L} have been limited to redshifts lower than $z \sim 1.5$ \citep[see][]{2004AJ....128.2677T, 2006ApJ...647L..21H, 2006MNRAS.370..198C, 2007MNRAS.376.1445C,2013ApJ...767...89M}. In contrast, photo-z surveys provide larger sample sizes and reach to higher redshifts, but they suffer from poorer redshift resolution. Figures \ref{fig:1} -- \ref{fig:22} show how the photo-z uncertainties distort the real galaxy environment from different viewpoints. The large-scale structures are clearly revealed in the case without photo-z error but become less prominent as the photo-z error increases. Despite this problem, there have been some attempts to measure galaxy environment for various studies using photo-z samples \citep{2007ApJS..172..284C, 2012ApJ...744...88Q, 2013ApJS..206....3S, 2014ApJ...782L...3C, 2016ApJ...817...97L}. Several works have provided viable methods that can be used to recover the density fields of galaxies from photo-z samples \citep{2015MNRAS.451..660E, 2016ApJ...817...97L, 2016A&A...585A.116M}. Moreover, \cite{2009MNRAS.394.1631A} and \cite{2012MNRAS.425.2099S} both demonstrated that the two-point correlation function of galaxies can also be successfully recovered from photometric samples, and they also discussed the influence of photo-z errors on their measurements. 

As several ongoing large sky surveys such as the Panoramic Survey Telescope \& Rapid Response System \citep[Pan-STARRS:][]{2008SPIE.7014E..12O, 2010SPIE.7733E..12K}, Dark Energy Survey \citep[DES:][]{2005astro.ph.10346T, 2006astro.ph..9591A}, Hyper Suprime-Cam Survey \citep[HSC:][]{2012SPIE.8446E..0ZM} and the upcoming Large Synoptic Survey Telescope \citep[LSST:][]{2009arXiv0912.0201L} will yield large galaxy samples with photometric redshift measurements, it is important to understand the potential and limitation of the photo-z method in the studies of galaxy evolution, especially the environmental effects. Can we use photo-z samples to measure environment reliably? What are the systematics in the environment measurement between spectral-z samples and photo-z samples? What is the optimal choice for density measurement that can reliably recover the underlying environments? These are the questions that we aim to answer. Particularly, we focus on the measurement of galaxy density field. We first study the difference between 3D real-space density and 2D projected density measurements by using mock galaxy catalogs. We adopt the Spearman's rank correlation coefficient \citep{10.2307/1412159}, $r_{s}$, as a measure of the correlation between the 2D and the 3D real-space density. The optimized parameters for the density measurement are obtained by maximizing $r_{s}$. We then use the results of optimized density measurements to show the dependence of galaxy properties on environments from the mock catalog. And finally, we apply our optimized scheme to the Pan-STARRS1 data and compare the results with the measurements by \cite{2006MNRAS.370..198C} who use the DEEP2 spectroscopic sample in the same field.

This paper is structured as follows. In Section \ref{sec:2.1} and \ref{sec:2.2} we describe the simulation and observational data used in our study. The environment measurements used in this study are introduced in Section \ref{sec:3}. In Section \ref{sec:4} we compare the 3D real-space density with 2D projected environment, and demonstrate how to optimize the choice of $N^{\rm th}$ nearest neighbor to improve 2D projected density measurement. In Section \ref{sec:5} we show the relation between galaxy environment and galaxy properties in the mock galaxy catalog to verify whether or not our optimized scheme is applicable. We discuss several possible factors that might limit our optimized scheme and apply it to observations in Section \ref{sec:6}. Finally we summarize our results in Section \ref{sec:7}. In this paper, we adopt the following cosmological parameters: $H_{0}=100h$ km $\rm s^{-1}$ $\rm Mpc^{-1}$, $h=0.73$, ${\Omega_0}=0.25$ and $\Omega{_\Lambda}=0.75$.

\section{DATA}
\subsection{Simulation Data}
\label{sec:2.1}

In this work, we use a theoretical mock galaxy catalog to understand the systematics in the local density estimates. The advantage of using a mock galaxy catalog compared to real spectroscopic survey data is that the real-space density can be directly measured and compared with the projected density. Moreover, the mock sample does not suffer from the incompleteness which often affects real observations. On the other hand, one needs to be cautious when interpreting the results since the properties of galaxies in the simulation may not be a perfect representation of the real Universe. 

The mock galaxy catalog used in this work is built based on Millennium simulation with $N = 2160^{3}$ in a box with volume = $500^{3}$ $h^{-3}$ $\rm Mpc^{3}$ from redshift $z = 127$ to the present day at $z = 0$ by adopting the following cosmological parameters: a baryon matter density ${\Omega_b}$ = 0.045, a total matter density ${\Omega_0}$ = 0.25, a dark energy density $\Omega{_\Lambda}$=0.75 and a Hubble constant $H_{0} = 100h$ km $\rm s^{-1}$ $\rm Mpc^{-1}$ where $h = 0.73$. These cosmological parameters match the first-year results of Wilkinson Microwave Anisotropy Probe \citep[WMAP:][]{2003ApJS..148..175S}. Galaxies are put into halos using the GALFORM semi-analytical model \citep{2000MNRAS.319..168C} which takes into account various galaxy formation processes including gas accretion and cooling, star formation in galactic disks and galaxy mergers. The mock catalog adopts the \cite{2012MNRAS.426.2142L} model which takes advantage of the extension to the treatment of star formation introduced into GALFORM in \cite{2011MNRAS.416.1566L} to populate galaxies, and is then assembled into a lightcone \citep{2013MNRAS.429..556M}. Further detailed information is given in \cite{2000MNRAS.319..168C}, \cite{2005Natur.435..629S}, \cite{2006MNRAS.370..645B}, \cite{2011MNRAS.416.1566L, 2012MNRAS.426.2142L} and \cite{2013MNRAS.429..556M}.

We constructed two types of mock catalogs that mimic the observed spectral-z and photo-z catalogs. The mock spectral-z catalog can be obtained from primitive simulation data which stores the intrinsic line-of-sight positions of galaxies. To generate mock photo-z catalogs, we perturb the position of galaxies along the line-of-sight direction by making a random shift which follows a Gaussian-distribution with standard deviation that matches the photo-z error in each case, in order to simulate cases with observed redshift uncertainties. The new redshift obtained can be viewed as the "observed redshift", and be used to compute the local density. Although the photometric redshift model adopted here is oversimplified as it does not take into account the effect of catastrophic redshift failures, this simplistic model allows us to understand the effect of redshift dispersion. Later in Section \ref{sec:5.4}, we consider more realistic situations in which the outlier effect is included. In this study we consider several photo-z cases with uncertainties of $\sigma_{\Delta_{z}/(1+z)}$ = 0.00, 0.02, 0.04 and 0.06. We restrict our environment study to the redshift range of $0.3 < z < 0.5$ for most of our analysis.  A central area of catalogs $\sim16$ square degrees is selected for our studies, containing $\sim1,900,000$ galaxies with $i < 25.8$. It is worth noting that the redshift which we use for computing the 3D overdensity with the Millennium mock spectroscopic catalogs refers to the intrinsic redshift of galaxies, which does not include the effect from peculiar velocity. Therefore, the results shown for the spectroscopic redshift sample may be too optimistic. However, since our main focus is to understand the performance of density recovery in the case of photometric errors, this mock spectroscopic catalog does provide a 'real' answer of the 3D density.

\subsection{Observation data}

\label{sec:2.2}

\subsubsection{Spectroscopic Observation}

The DEEP2 Galaxy Survey \citep{2003SPIE.4834..161D, 2013ApJS..208....5N} was designed to study the galaxy population and large-scale structure at $z\sim1$. It uses the Keck II telescopes with the DEIMOS spectrograph \citep{2003SPIE.4841.1657F} and covers $\sim 3.5$ square degrees of the sky with measured spectra and has targeted $\sim$ 60,000 galaxies down to a limiting magnitude of $R_{\rm AB} < 24.1$. About $\sim60\%$ of the galaxies are sampled over the redshift interval $0.2 < z < 1.4$. The overall redshift success rate is about $\sim70\%$. DEEP2 comprises four widely separated fields. One of the DEEP2 fields, the Extended Groth Strip (EGS), is enclosed by the Pan-STARRS1 Medium Deep Survey Field (MD07). In this study we match the DEEP2 spectroscopic redshift catalog to the Pan-STARRS1 MD07 catalog. For galaxies that are common in the two catalogs, we compare the local density measurements computed using the DEEP2 spectral-z and Pan-STARRS1 photo-z respectively (see Section \ref{sec:6}) .

\subsubsection{Photometric Observation}

Pan-STARRS1 (hereafter PS1) is a 1.8 meter telescope equipped with a CCD digital camera with 1.4 billion pixels and 3-degree field of view, located on the summit of Haleakala on Maui in the Hawaii Islands \citep{2008SPIE.7014E..12O, 2010SPIE.7733E..12K}. The PS1 observations are obtained in a set of five broadband filters, which we have designated as $g_{p1}, r_{p1}, i_{p1}, z_{p1}$ and $y_{p1}$. There are two major components of the PS1 survey which started observations in 2010: the 3$\pi$ survey and the Medium Deep Survey (MDS) which comprises ten fields spread across the sky. One of the MDS fields, namely MD07, is chosen for this study because it overlaps with the EGS field which has the spectroscopic from the DEEP2 survey, and enables a direct comparison with the environment measurements using the DEEP2 spectral-z sample \citep{2006MNRAS.370..198C}. Photo-z in MD07 are computed by running the EAZY code \citep{2008ApJ...686.1503B} on PS1 5-band photometry plus the $u^{*}$-band data taken by Eugene Magnier et al. with CFHT MEGACAM as part of the PS1 efforts. Comparisons against the DEEP2 spectroscopic redshifts \citep{2013ApJS..208....5N} show that the PS1 photo-z reaches an uncertainty of 0.05 with outlier rate of 7\% down to $r_{p1} < 24.1$.  More details on the data processing and photo-z characteristics in the PS1 MD07 sample are given in \cite{2014ApJ...782...33L}, and Foucaud et al. (in preparation). 

\section{Galaxy Environment Measurements}
\label{sec:3}

In this study, we adopt the $N^{\rm th}$ Nearest Neighbor method to quantify galaxy environment. This method defines the local density of each galaxy using the distance to the $N^{\rm th}$ nearest neighbor galaxy. In other words, whether the galaxy is located in an over-dense or under-dense environment depends on how far it is to its $N^{\rm th}$ nearest galaxy. In simulations, the cosmological redshift reflects the "real" distance along the line of sight, enabling the environment measurement to be evaluated by using 3D $N^{\rm th}$ Nearest Neighbor method. On the other hand, observationally, the local density is often estimated by using a projected method as the measured redshift is a combination of both the cosmological distance and the peculiar velocity.

\begin{figure}
\epsscale{1.0}
\plotone{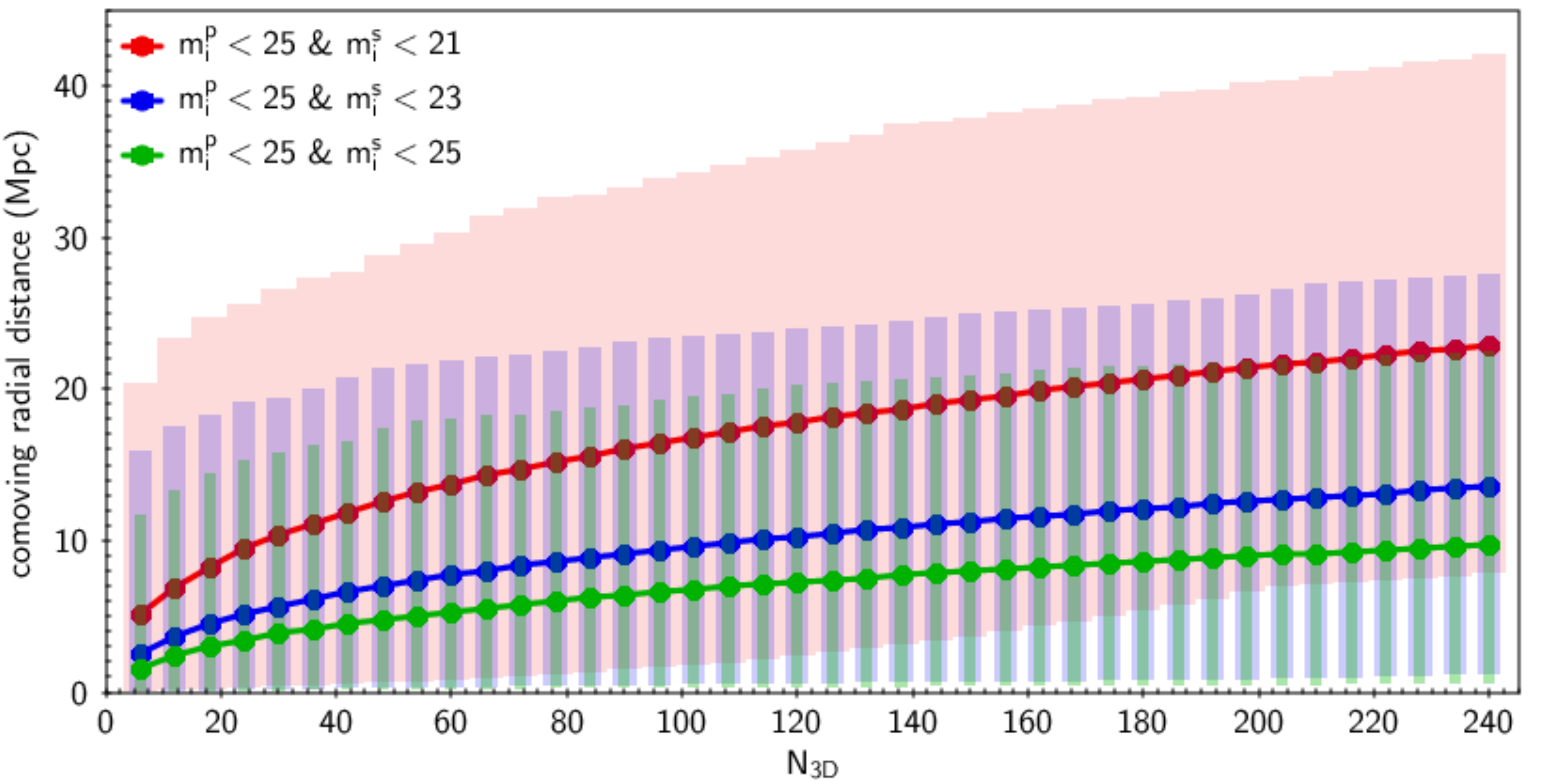}
\caption{The median $N^{\rm th}$ nearest-neighbor distance as a function of $N_{\rm 3D}$ for various choices of the magnitude limit of the secondary sample ($m_{i}^{s}$). The shadows of different colors show the range between the minimum and maximum $N^{\rm th}$ nearest-neighbor distance. \label{fig:29} }
\end{figure}

Here we define two sets of galaxy samples: the primary and the secondary sample. The primary sample contains galaxies brighter than a particular magnitude ($m_{i}^{p}$), and is used when presenting the results. The secondary sample refers to galaxies used in the search for neighbors, and is restricted to those galaxies which are brighter than a particular magnitude ($m_{i}^{s}$). The limiting magnitude of the secondary sample is particularly important because it sets the galaxy number densities in the calculation of density field. Figure \ref{fig:29} shows the median distances to the 3D $N^{\rm th}$ nearest neighbor with various choices of $m_{i}^{s}$. The coloured shadows show the range between the minimum and maximum distances to the 3D $N^{\rm th}$ nearest neighbor. There are several parameters that should be considered in the $N^{\rm th}$ Nearest Neighbor method: (1) the choice of the $N^{\rm th}$ Neighbor, which represents the scale of environment, (2) the magnitude limit of the primary sample ($m_{i}^{p}$), (3) the magnitude limit of the secondary sample ($m_{i}^{s}$), and (4) the velocity window ($V_{\rm cut}$) that defines the redshift boundaries of the neighbors considered in the 2D projected method. These parameters should be adjusted according to different science goals and galaxy samples. One of the goals of this work is to provide an empirical framework which determines these parameters by calculating galaxy densities with different combinations of parameters in order to understand the influence of the parameters.

\subsection{2D Projected $N^{\rm th}$ Nearest Neighbor Galaxy Environment}

\label{sec:3.1}

In the 2D projected method, the local density of each galaxy is computed as the surface density averaged over the area enclosed by the $N^{\rm th}$ closest galaxy within the velocity interval $V_{\rm cut}$: 

\begin{equation}
	\Sigma_n = \frac{n+1}{\pi r^2_n},
\end{equation}
where n is the $N^{\rm th}$ closest galaxy for each reference galaxy and $r_{n}$ is the distance from the reference galaxy to the $N^{\rm th}$ closest galaxy on a 2D surface. There is no simple way to determine the choice of velocity window $V_{\rm cut}$ in the 2D nearest-neighbor method. In principle, it is not meaningful to adopt a $V_{\rm cut}$ that is too small compared to the redshift uncertainty of the data. Conversely, adopting a large velocity cut enlarges projection effect, which leads to greater errors in the density measurement. For instance, the separation between two galaxies that are close in the projected plane may actually be widely separated in the third dimension, and vice versa \citep{2012MNRAS.419.2670M}. Previous studies have utilized the velocity interval $V_{\rm cut}$ of a value close to the distance uncertainties in the line-of-sight direction, as they found that the density estimate does not significantly vary when changing $V_{\rm cut}$ around this value. We will further test this approach using the mock catalog in Section \ref{sec:4.1}.

\subsection{3D $N^{\rm th}$ Nearest Neighbor Galaxy Environment}

\label{sec:3.2}

The galaxy environment defined by the 3D $N^{\rm th}$ Nearest Neighbor method is similar to that defined by the 2D projected $N^{\rm th}$ Neighbor except that the projected circular area is replaced by the enclosed-spherical volume. The volume density of galaxies is evaluated using 3D $N^{\rm th}$ Nearest Neighbor method as:

\begin{equation}
	\rho_{n} = \frac{n+1}{(4/3)\pi r^3_n},
\end{equation}
where n is the $N^{\rm th}$ closest galaxy of each reference galaxy and $r_{n}$ is the distance from the reference galaxy to the $N^{\rm th}$ closest galaxy in the three-dimensional space. We compute the real-space density using the 3D $N^{\rm th}$ Nearest Neighbor method of the simulation where information about the three-dimensional positions of galaxies is known. We treat the 3D density as the "true" density to be compared with the 2D density to quantify how well the real-space density can be recovered by the 2D projected density under various conditions. We note that practically the 3D density is rarely used even in a spectroscopic redshift sample, because the observed 'redshift' includes contributions from both the Hubble flow and the peculiar velocity of galaxies, which is not possible to differentiate observationally. 

Finally, in order to contrast the most-dense environments with the least-dense environments, we convert the initial primordial density into an overdensity. The overdensity is conventionally defined as the initial primordial density divided by the median density as follows:  

\begin{equation}
	1+\delta_{n} = \frac{D_{i}}{D_{Mdn} },
\end{equation}
where $D_{i}$ is the measured density of a galaxy (i.e., $D_{i} \in \{\rho_{n}, \Sigma_{n}\}$), and $D_{Mdn}$ is the median density computed by counting galaxies within a bin of $\Delta$z = 0.04. The term $1+\delta_{n}$ is the so-called overdensity, and $\delta_{n}$ can be $\delta_{n}^{3D}$ or $\delta_{n}^{2D}$, depending on $D_{i}$ = $\rho_{n}$ or $\Sigma_{n}$.

\section{Quantifying differences between Environment Measurements}
\label{sec:4}

In this section, we compare the 2D projected density obtained under various conditions to the 3D real-space density. To quantify their differences, we adopt the Spearman's rank correlation coefficient, $r_{s}$, which is commonly used to measure the strength of a relationship between two ranked variables \citep{10.2307/1412159, 2014arXiv1411.3816C}. The Spearman's rank correlation coefficient is defined as: 

\begin{equation}
	r_{s} = 1 - \frac{6 \sum d^{2}_{i}}{s(s^{2}-1)}\label{formula:4},
\end{equation}
where $d_{i}$ is the difference between the ranks of the two variables and $s$ is the sample size. A coefficient $r_{s}$ = 0 corresponds to no correlation between two variables, while $r_{s}$ = 1 (-1) corresponds to a perfect-positive (perfect-negative) correlation. In our analysis, we first measure the 2D projected density $\Sigma_{n}$ as well as the 3D real-space density $\rho_{n}$ for each galaxy, and then convert all the density measurements into an overdensity as defined in the last section. After ranking the 3D real-space overdensity and 2D projected overdensity, we compute the difference $d_{i}$ between the ranks of the two overdensities, and then use Formula \ref{formula:4} to calculate Spearman's rank correlation coefficient $r_{s}$.

\begin{figure}
\epsscale{1.0}
\plotone{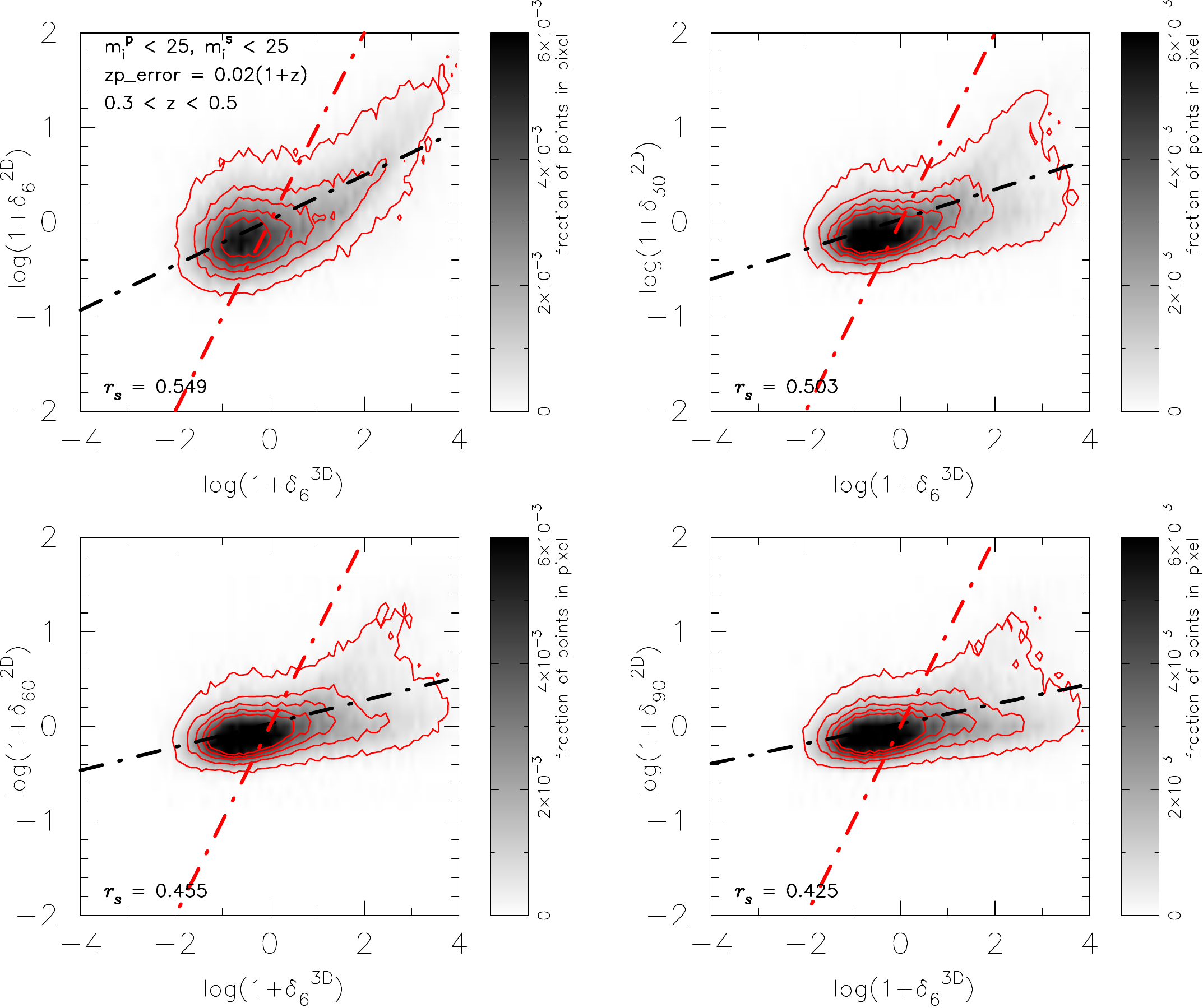}
\caption{The scatter plot of the 3D real-space overdensity, $1+\delta_{6}^{3D}$, versus 2D projected overdensity: $1+\delta_{6}^{2D}$ (upper-left panel), $1+\delta_{30}^{2D}$ (upper-right panel), $1+\delta_{60}^{2D}$ (lower-left panel) and $1+\delta_{90}^{2D}$ (lower-right panel) in log-scale. The numbers printed in the bottom-left of each panel indicate the $r_{s}$ coefficient. The black dash-dot lines represent the best fit to the data points, the red dash-dot lines represent the one-to-one relation and the contours show the regions of constant galaxy number. \label{fig:2} }
\end{figure}

Figure \ref{fig:2} is an example showing the correlation between the 3D and 2D measurements using $N_{\rm 2D}$ = 6, 30, 60 and 90 respectively, and how the $r_{s}$ coefficient changes with different choices of $N_{\rm 2D}$. 

\subsection{3D Galaxy Environment vs. 2D Projected Galaxy Environment with Different Parameters}

\label{sec:4.1}

\begin{figure}
\epsscale{1.0}
\plotone{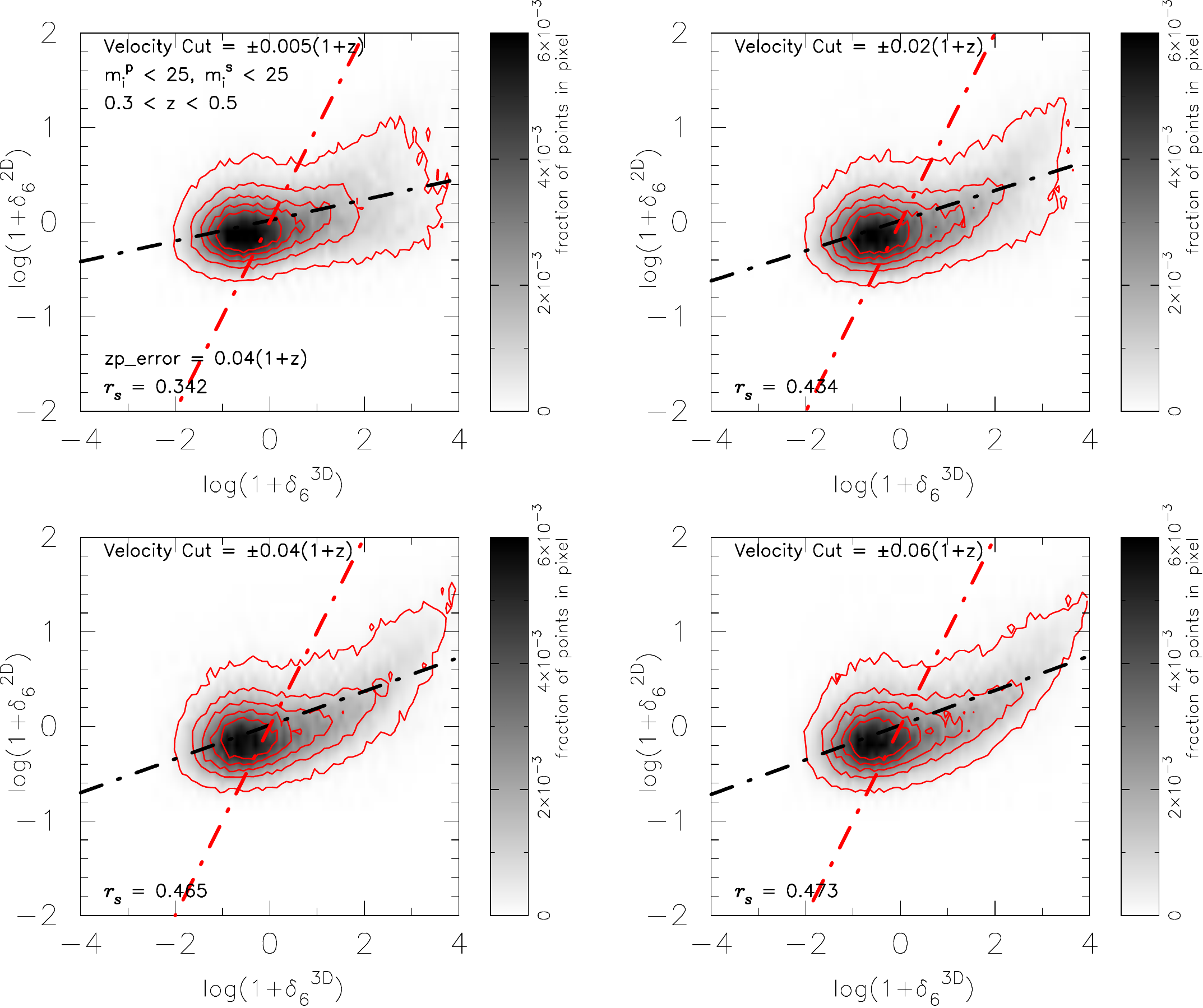}
\caption{The scatter plot of the 3D real-space overdensity, $1+\delta_{6}^{3D}$, versus 2D projected overdensity, $1+\delta_{6}^{2D}$, with various velocity cuts: $V_{\rm cut}$ = $\pm0.005(1+z)$, $\pm0.02(1+z)$, $\pm0.04(1+z)$ and $\pm0.06(1+z)$ over the redshift interval $0.3 < z < 0.5$. All cases are considered by using galaxy samples with photo-z error = $0.04(1+z)$, $m_{i}^{s} < 25$ and $m_{i}^{p} < 25$. The numbers printed in the bottom-left of each panel indicate the $r_{s}$ coefficient. The black dash-dot lines represent the best fit to the data points, the red dash-dot lines represent the one-to-one relation and the contours show the regions of constant galaxy number. \label{fig:3} }
\end{figure}

We first probe the effect of the velocity interval $V_{\rm cut}$ on the 2D projected density measurement. We restrict the sample with $m_{i}^{p} < 25$ and $m_{i}^{s} < 25$ when calculating 2D projected overdensity, $1+\delta_{6}^{2D}$, and 3D real-space overdensity, $1+\delta_{6}^{3D}$. Figure \ref{fig:3} shows the scatter plots of the 2D projected overdensity versus the real-space overdensity in log scale using the 6$^{\rm th}$ nearest neighbor. Here we consider the following four different choices of $V_{\rm cut}$ in the 2D projected measurement for a galaxy sample with photo-z error = $0.04(1+z)$: $\pm0.005(1+z)$, $\pm0.02(1+z)$, $\pm0.04(1+z)$ and $\pm0.06(1+z)$. It can be seen that the difference in $r_{s}$ among the four cases of $V_{\rm cut}$ is not significant when the size of velocity interval is close to the photo-z error. For example, the $r_{s}$ of 0.434 is obtained in the case of $V_{\rm cut}$ = $\pm0.02(1+z)$ (upper-right panel of Figure \ref{fig:3}), while the value of $r_{s}$ increases to 0.473 in the case of $V_{\rm cut}$ = $\pm0.06(1+z)$ (bottom-right panel of Figure \ref{fig:3}). The difference in $r_{s}$ is small ($\sim$0.039) between these two cases even though the $V_{\rm cut}$ differs by a factor of 3, which confirms the finding in previous studies that the 2D projected measurement is not sensitive to $V_{\rm cut}$ when $V_{\rm cut}$ is comparable to the photo-z uncertainty \citep{2009ApJ...690.1883G, 2012MNRAS.419.2670M}. We further repeat similar exercises using samples with a larger redshift uncertainty up to $0.08(1+z)$ and at higher redshifts ($0.6 < z < 0.8$) and find that this conclusion still holds. Therefore throughout this work, we set $V_{\rm cut}$ to be the typical photo-z uncertainty of the galaxy sample.

\begin{figure}
\epsscale{1.0}
\plotone{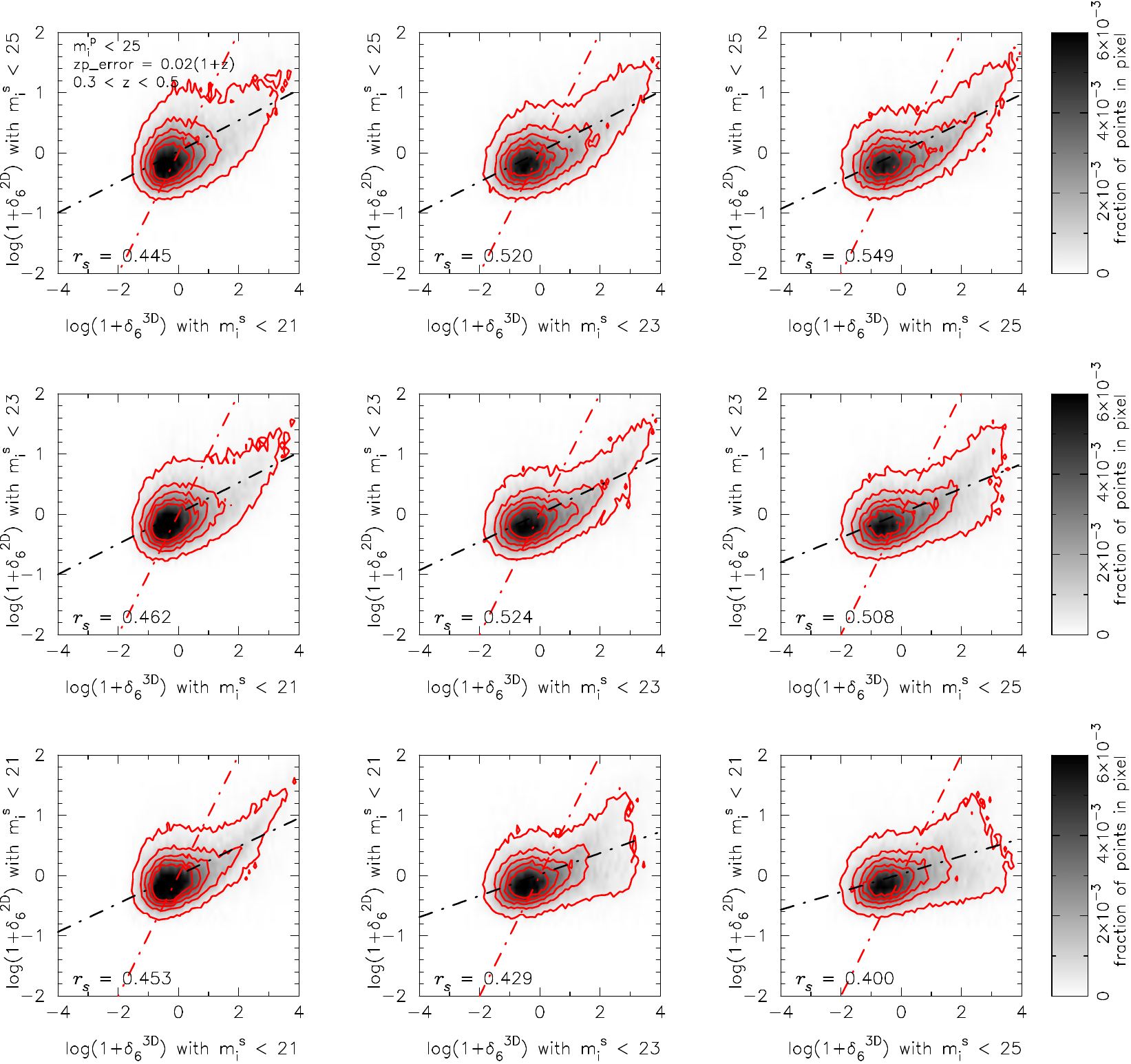}
\caption{The scatter plot of the 3D real-space overdensity, $1+\delta_{6}^{3D}$, versus 2D projected overdensity, $1+\delta_{6}^{2D}$, with various secondary magnitude limit in 2D measurements (series of row panels, from left to right: $m_{i}^{s} < 21$, $m_{i}^{s} < 23$ and $m_{i}^{s} < 25$) and 3D measurements (series of column panels, from bottom to top: $m_{i}^{s} < 21$, $m_{i}^{s} < 23$ and $m_{i}^{s} < 25$) over the redshift interval $0.3 < z < 0.5$. All cases are considered by using galaxy samples with photo-z error = $0.02(1+z)$, $V_{\rm cut}$ = $\pm0.02(1+z)$ and $m_{i}^{p} < 25$. The numbers printed in the bottom-left of each panel indicate the $r_{s}$ coefficient. The black dash-dot lines represent the best fit to the data points, the red dash-dot lines represent the one-to-one relation and the contours show the regions of constant galaxy number. \label{fig:4} }
\end{figure}

Next we consider the effect of the secondary magnitude limit employed on the galaxy sample when searching for neighbors. A brighter (fainter) $m_{i}^{s}$ probes a larger (smaller) scale of environment for a fixed $N^{\rm th}$ nearest neighbor. It is therefore expected that the correlation between the 2D projected and 3D real-space environment could depend on the choice of $m_{i}^{s}$. Again we select the sample with photo-z error = $0.02(1+z)$ and set $V_{\rm cut}$ = $\pm0.02(1+z)$ for the reason given above. Figure \ref{fig:4} shows the scatter plots of the 2D projected overdensity versus real-space overdensity using the 6$^{\rm th}$ nearest neighbor, and both 2D and 3D environments are measured using $m_{i}^{s} < 21$, $m_{i}^{s} < 23$ and $m_{i}^{s} < 25$ respectively. As it can be seen, the largest $r_{s}$ is obtained when the $m_{i}^{s}$ in 2D measurement is equal to the $m_{i}^{s}$ in 3D measurement. Furthermore, for identical $m_{i}^{s}$ used in the 2D and 3D measurements, the environments measured by using fainter $m_{i}^{s}$ (and hence smaller scales) have better correlation than those measured using a brighter $m_{i}^{s}$. This is consistent with the results from \cite{2013MNRAS.433.3314S}, which also shows that for samples with photo-z error, the 2D projected environments have a weaker correlation with 3D real-space environments on larger scales.

\begin{figure}
\epsscale{1.0}
\plotone{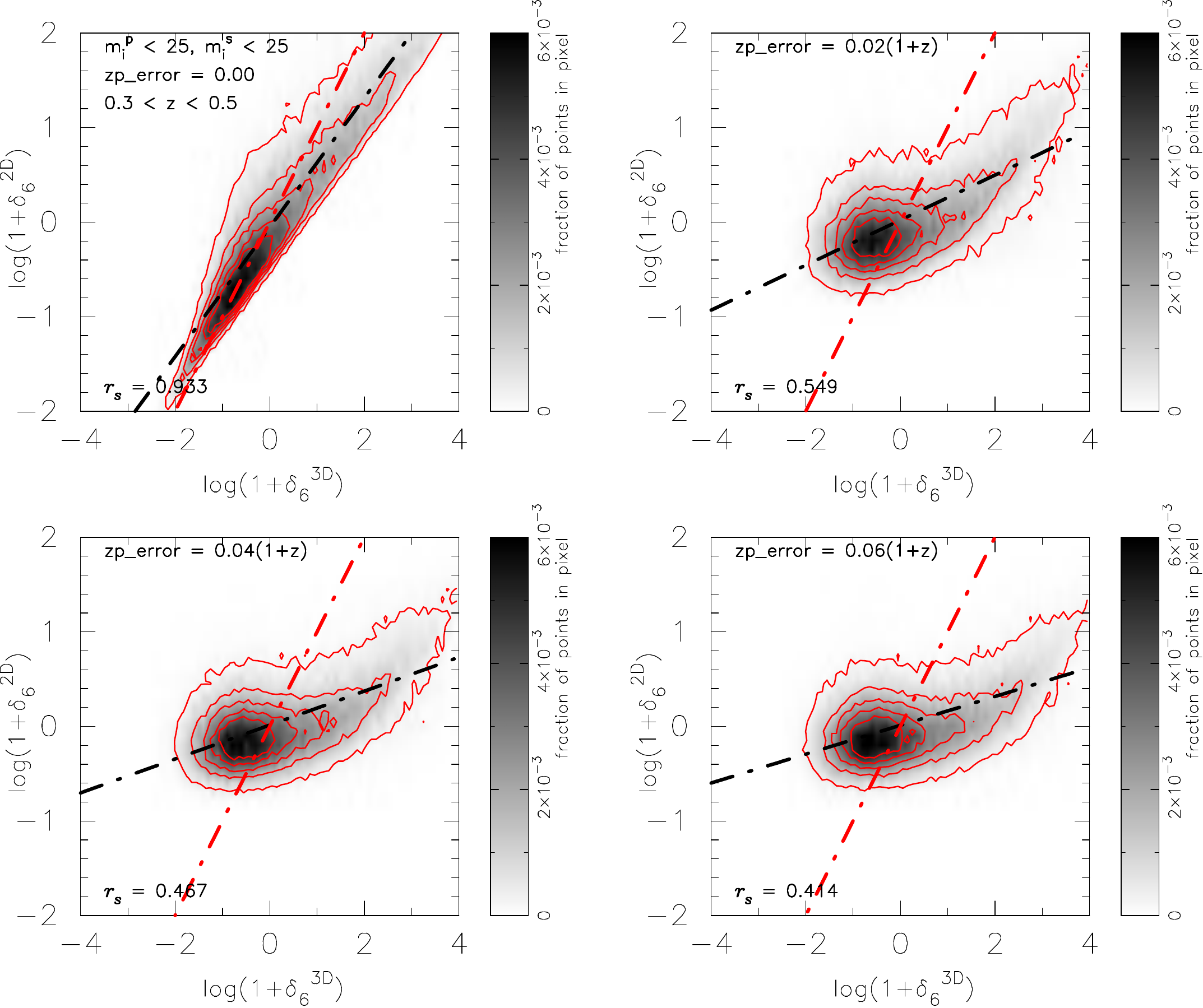}
\caption{The scatter plot of the 3D real-space overdensity, $1+\delta_{6}^{3D}$ versus 2D projected overdensity, $1+\delta_{6}^{2D}$, with different photo-z errors over the redshift interval $0.3 < z < 0.5$. All cases are considered by using galaxy samples with photo-z error = 0.00, $0.02(1+z)$, $0.04(1+z)$ and $0.06(1+z)$ and $V_{\rm cut}$ = $\pm0.001(1+z)$, $\pm0.02(1+z)$, $\pm0.04(1+z)$ and $\pm0.06(1+z)$ respectively. The primary and secondary magnitude limits are $m_{i}^{p} < 25$ and $m_{i}^{s} < 25$. The numbers printed in the bottom-left of each panel indicate the $r_{s}$ coefficient. The black dash-dot lines represent the best fit to the data points, the red dash-dot lines represent the one-to-one relation and the contours show the regions of constant galaxy number. \label{fig:5} }
\end{figure}

Finally, Figure \ref{fig:5} shows the 2D versus 3D scatter plots in the density measurement to understand how the galaxy environment is affected by the photo-z errors. Here we consider $m_{i}^{p} < 25$ and $m_{i}^{s} < 25$, and vary photo-z error from 0, $0.02(1+z)$, $0.04(1+z)$ to $0.06(1+z)$. $V_{\rm cut}$ is correspondingly set to be $\pm0.001(1+z)$, $\pm0.02(1+z)$, $\pm0.04(1+z)$ and $\pm0.06(1+z)$ respectively. It is worth noting that even under the perfect situation where the redshift error is zero, the correlation is still not perfect owing to the projection effect. The correlation between 3D real-space and 2D projected environments become gradually worse when the photo-z error increases. However, there still exists some correlation especially for galaxies located in high density regions, while the environment in lower and intermediate densities is less distinguishable. This is consistent with the result from \cite{2007ApJS..172..284C}, who also shows that the galaxy environment is difficult to measure for galaxies located in low dense regions when a redshift error is present.

\subsection{Optimizing 2D Environment Parameters}

\label{sec:4.2}

\begin{figure}
\epsscale{1.0}
\plotone{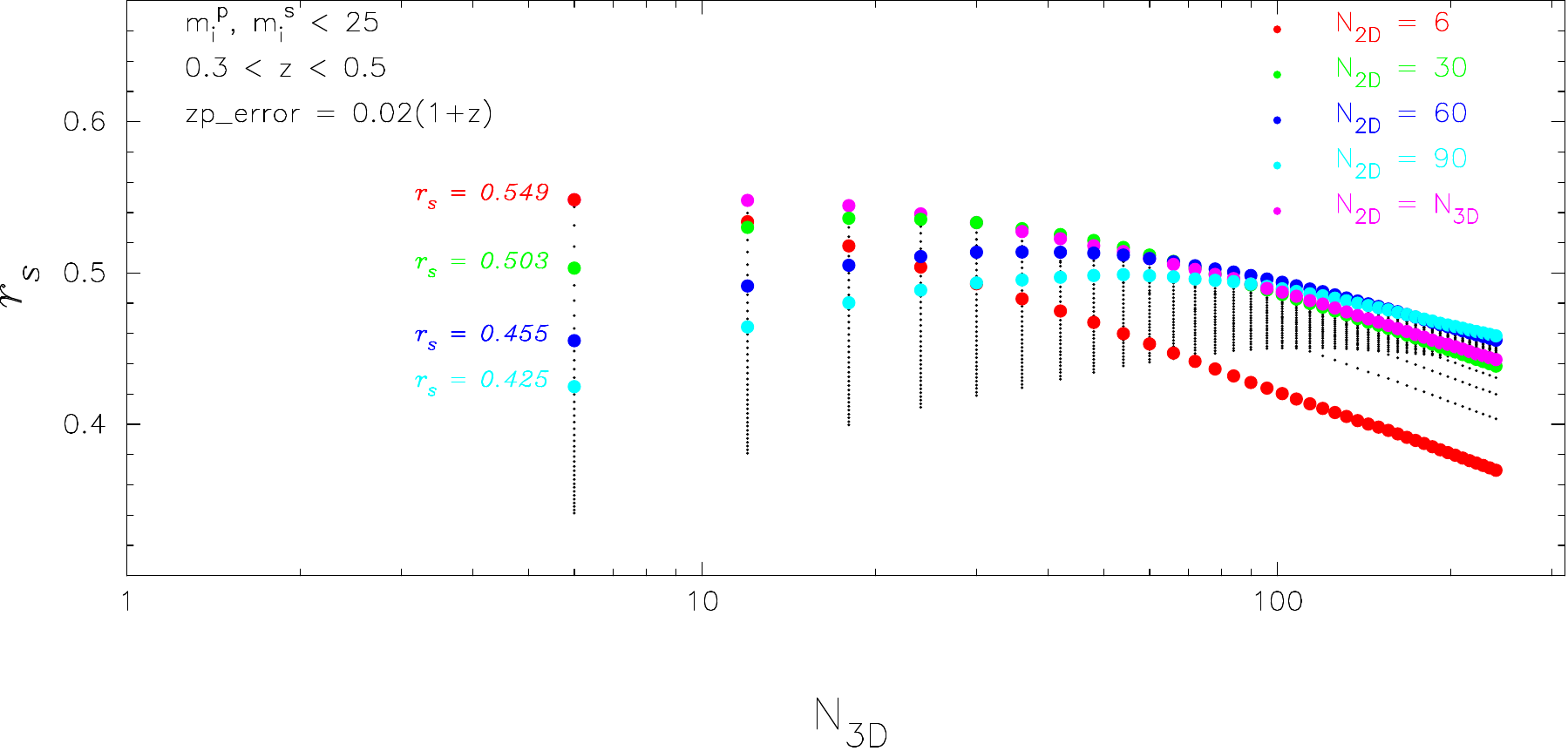}
\caption{The black dots show the $r_{s}$ coefficients calculated by fitting the real-space and 2D projected environments for different choices of $N_{\rm 2D}$ as a function of $N_{\rm 3D}$ from mock galaxy catalogs. The $r_{s}$ for the choice of $N_{\rm 2D}$ = 6, 30, 60, 90 and $N_{\rm 3D}$ are shown in red, green, blue, cyan and magenta dots, respectively. The values of $r_{s}$ corresponding to the four cases (0.549, 0.503, 0.455, 0.425) shown in Figure \ref{fig:2} are also marked. \label{fig:6} }
\end{figure}

So far we have compared 3D real-space environments with various 2D projected environments to show their correlation and we have adopted $r_{s}$ to quantify the goodness of the correlation. We now expand this to construct an optimization scheme to determine the value of $N_{\rm 2D}$ which gives the best correlation (largest $r_{s}$) between the 2D and 3D environments. Figure \ref{fig:6} shows the $r_{s}$ calculated by fitting 2D projected and 3D real-space environment with various choice of $N_{\rm 2D}$ and $N_{\rm 3D}$. The red, green, blue, cyan and magenta dots are for the cases where the 2D projected environments are calculated using $N_{\rm 2D}$ = 6, 30, 60, 90 and $N_{\rm 3D}$ respectively. We also mark the four cases of Figure \ref{fig:2}, $r_{s}$ = 0.425, 0.455, 0.503 and 0.549 respectively, to demonstrate how $r_{s}$ varies with different choices of $N_{\rm 2D}$. In the following analysis, we only show the optimized choice of $N_{\rm 2D}$ corresponding to the case which yields the largest $r_{s}$, as a function of $N_{\rm 3D}$.

\begin{figure}
\epsscale{1.0}
\plotone{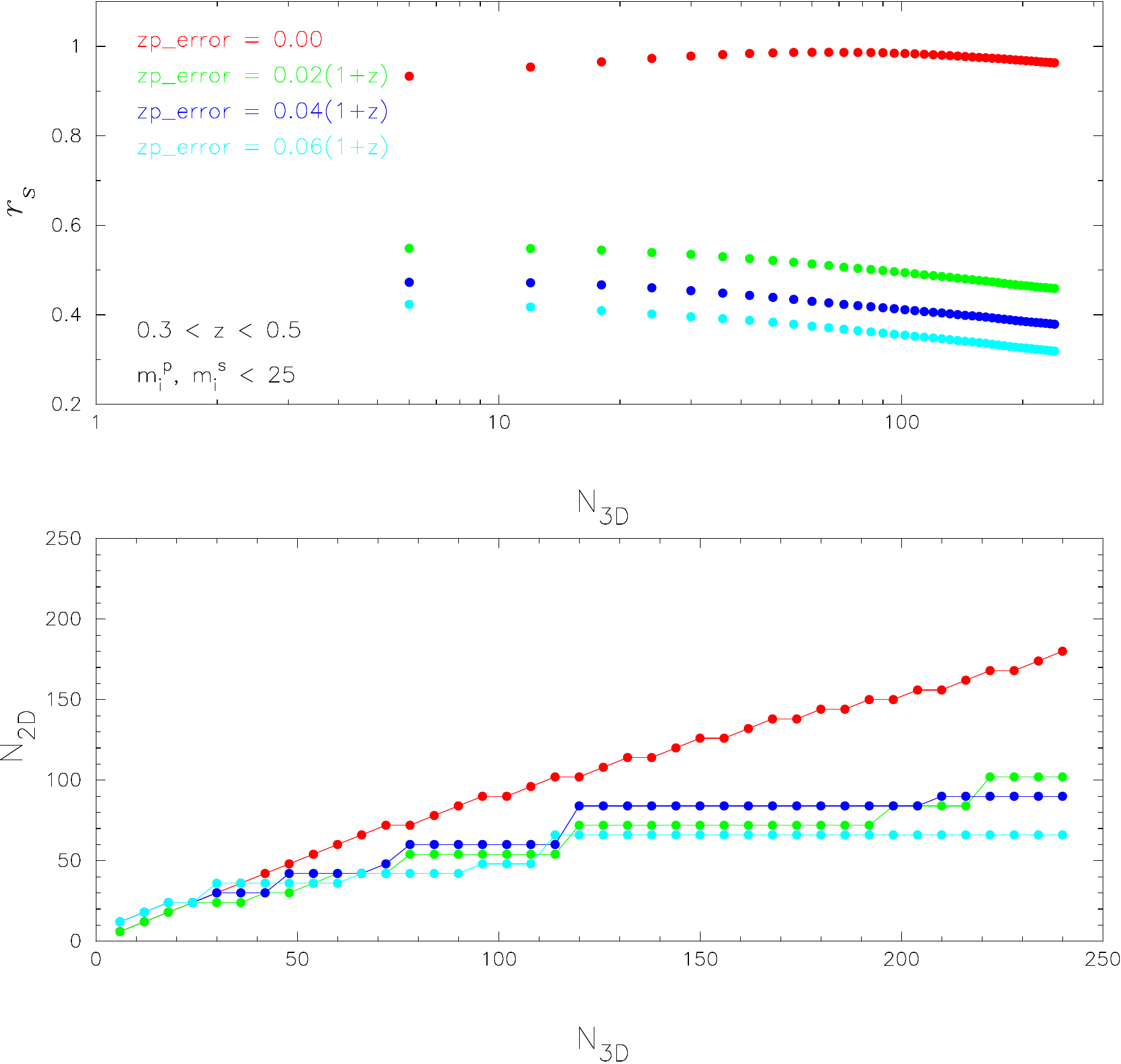}
\caption{The largest $r_{s}$ (upper-panel) obtained by varying $N_{\rm 2D}$ and the corresponding choice of $N_{\rm 2D}$ (lower-panel) that yields the best correlation between the real-space and 2D projected environments for different choices of $N_{\rm 2D}$ as a function of $N_{\rm 3D}$ from mock galaxy catalogs. Different color-dots correspond to samples with different photo-z errors: 0.0, $0.02(1+z)$, $0.04(1+z)$ and $0.06(1+z)$. \label{fig:7} }
\end{figure}

Figure \ref{fig:7} shows the largest $r_{s}$ (upper-panel) and corresponding choice of $N_{\rm 2D}$ (bottom-panel) that yields the best correlation between 2D projected and 3D real-space environments for different choices of $N_{\rm 2D}$, as a function of $N_{\rm 3D}$ from mock galaxy catalogs. The red, green, blue and cyan dots are for samples with different photo-z errors: 0.00, $0.02(1+z)$, $0.04(1+z)$, $0.06(1+z)$ respectively. As expected, the densities are relatively easier to recover for samples with lower photo-z errors than those with higher photo-z errors. The $r_{s}$ obtained for the case without photo-z error (red dots) are greater than 0.9, meaning that the optimized 2D projected environments are strongly correlated with the 3D real-space environments. However, for the samples contaminated by photo-z errors (green, blue and cyan dots), the recovering performance becomes gradually worse as the photo-z error increases. In addition, the correlation between 2D projected and 3D real-space environments does not depend only on the redshift accuracy, but is also scale dependent. For example, at a given photo-z error, $r_{s}$ decreases with $N_{\rm 3D}$, which suggests that small-scale environments are more easier to recover. A possible explanation is that the 2D projected environments calculated by using photo-z samples might include more contaminations when we probe the larger scale of environments.

One interesting feature in the bottom panel of Figure \ref{fig:7} is that the best choices of $N_{\rm 2D}$ are in general not equal to $N_{\rm 3D}$, but only half of $N_{\rm 3D}$, for producing the largest $r_{s}$ except for the case of error-free. However, we note that the relation between the optimized $N_{\rm 2D}$ and $N_{\rm 3D}$ depends on the choices of redshift interval. Detailed discussions are given in Appendix \ref{sec:8.1}.

\begin{figure}
\epsscale{1.0}
\plotone{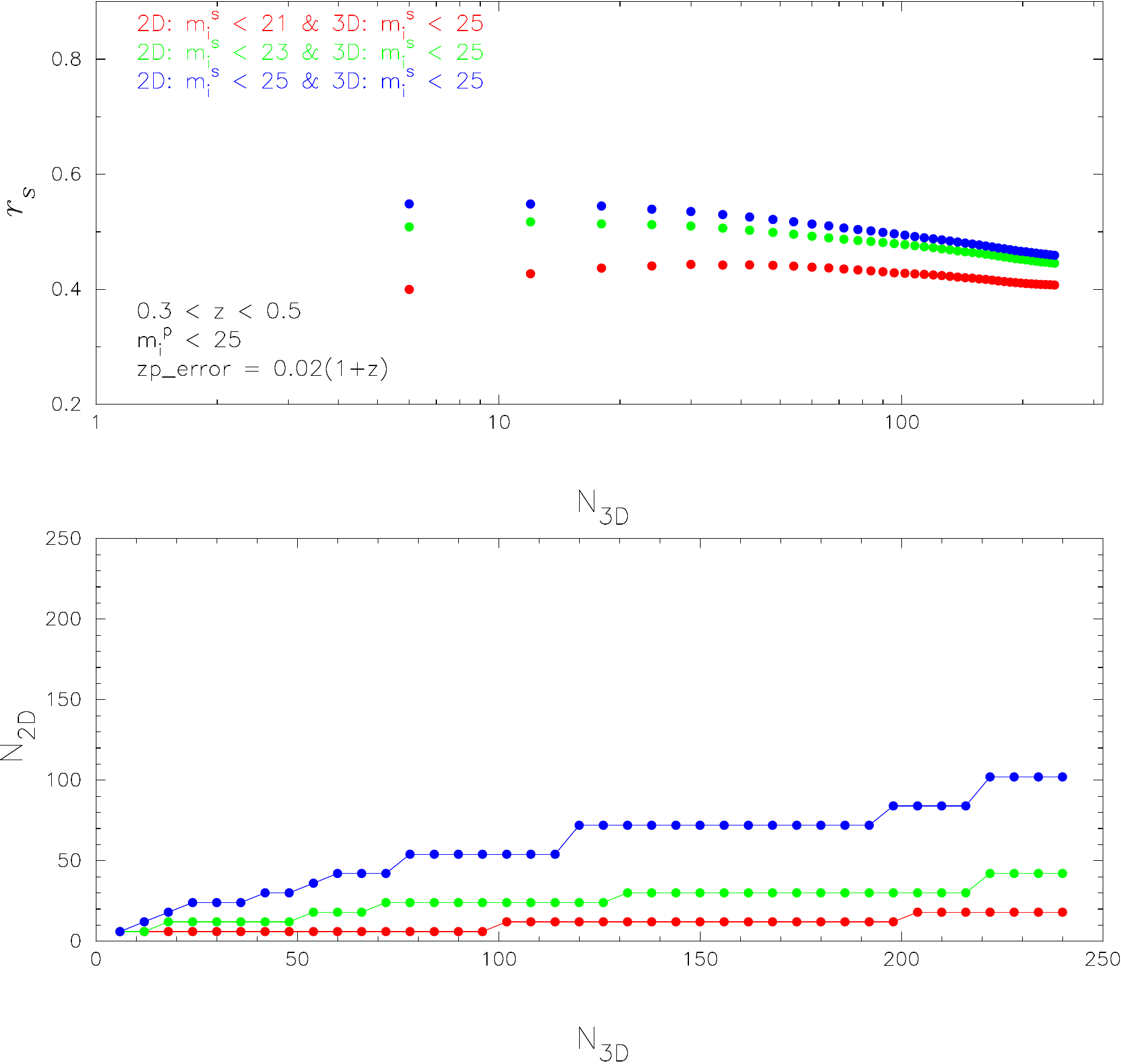}
\caption{The largest $r_{s}$ (upper-panel) obtained by varying $N_{\rm 2D}$ and the corresponding choice of $N_{\rm 2D}$ (lower-panel) that yields the best correlation between the real-space and 2D projected environments for different choices of $N_{\rm 2D}$ as a function of $N_{\rm 3D}$ from mock galaxy catalogs. Here the secondary magnitude limit in 3D measurement is set to $m_{i}^{s} = 25$. Different color-dots correspond to samples with different secondary magnitude limits in 2D measurements: $m_{i}^{s} < 21$, $m_{i}^{s} < 23$ and $m_{i}^{s} < 25$. \label{fig:8_1} }
\end{figure}

\begin{figure}
\epsscale{1.0}
\plotone{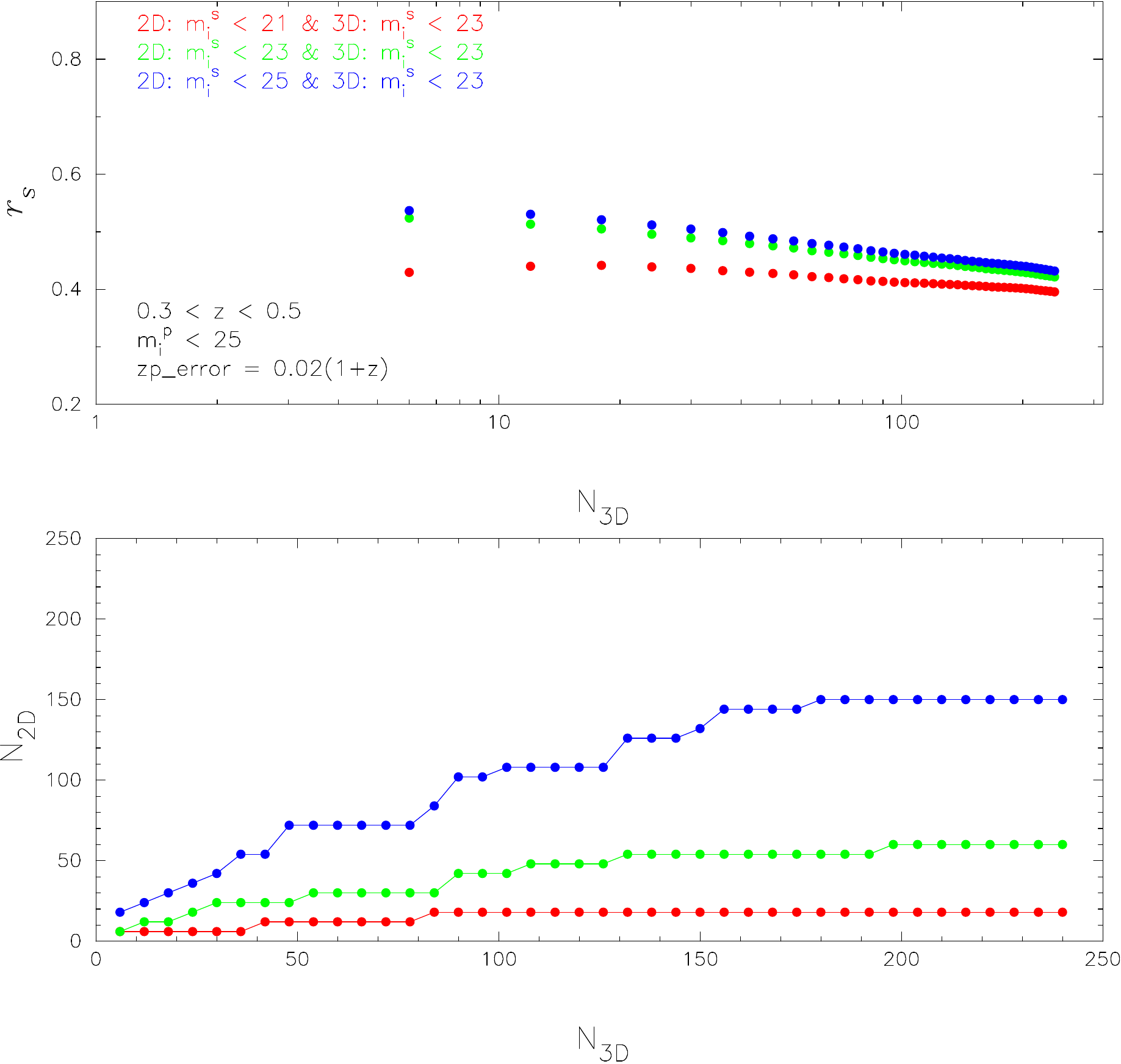}
\caption{Similar to Figure \ref{fig:8_1} but the secondary magnitude limit in 3D measurement is set to $m_{i}^{s} = 23$. \label{fig:8_2} }
\end{figure}

\begin{figure}
\epsscale{1.0}
\plotone{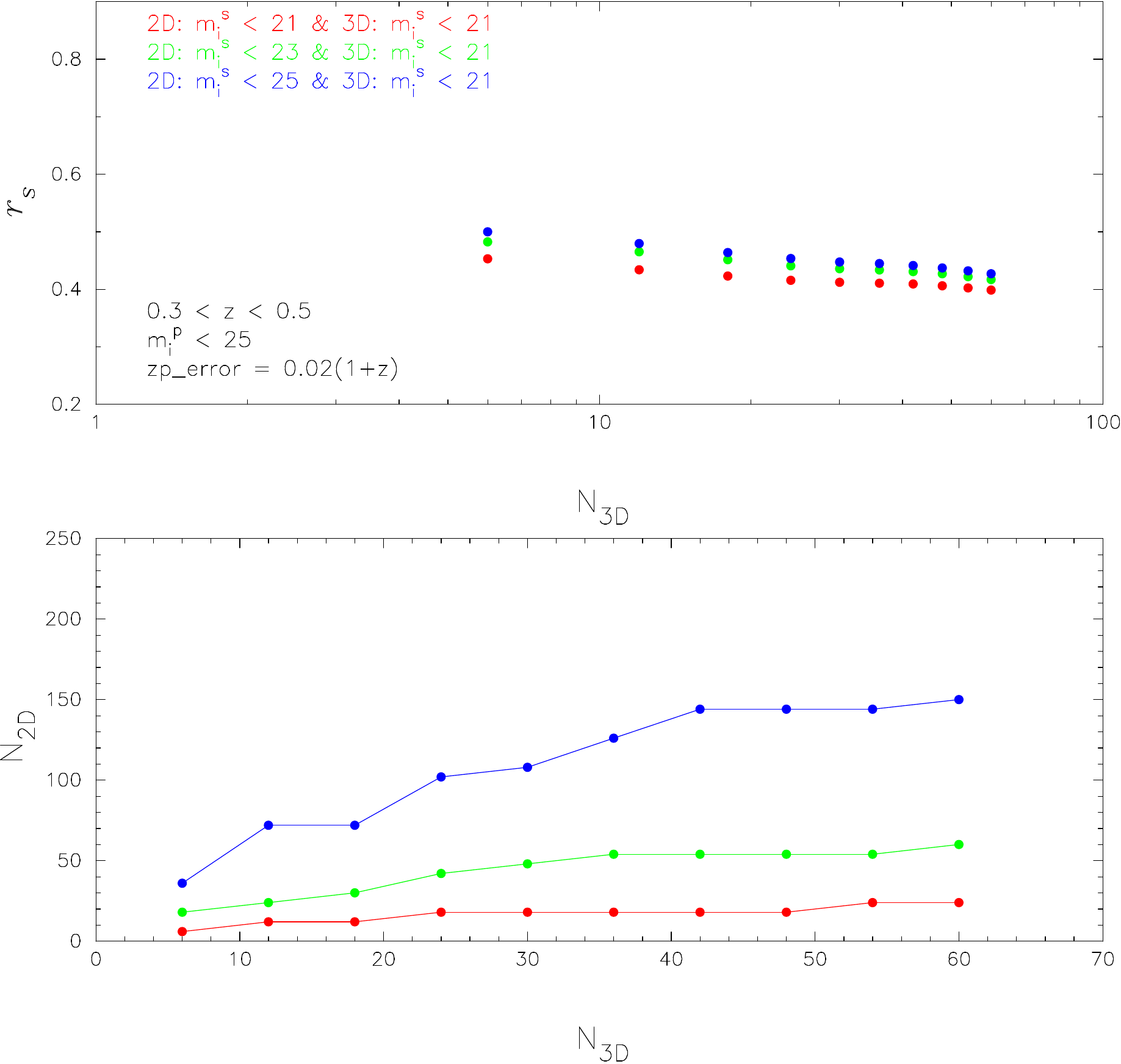}
\caption{Similar to Figure \ref{fig:8_1} but the secondary magnitude limit in 3D measurement is set to $m_{i}^{s} = 21$. \label{fig:8_3} }
\end{figure}

As the value of density measurements also depends on the choice of the secondary magnitude limit, it is interesting to see how the correlation between 2D and 3D measurements change by varying the secondary magnitude limits in both 3D and 2D environment measurements. Figures \ref{fig:8_1} -- \ref{fig:8_3} show the largest $r_{s}$ as a function of $N_{\rm 3D}$ for different 3D secondary magnitude limits respectively. In each figure, the red, green and blue dots are for samples with secondary magnitude limits corresponding to $m_{i}^{s} < 21$, $m_{i}^{s} < 23$, and $m_{i}^{s} < 25$ in 2D environment measurement respectively, at a fixed secondary magnitude limit in 3D local density and a fixed photo-z error = $0.02(1+z)$. Our results show that when the 3D environment is defined using brighter secondary magnitude limits, there is no significant difference in the recovering performance among different choices of 2D secondary magnitude limit that are fainter than the 3D secondary magnitude limit. On the other hand, adopting a 2D secondary magnitude limit that is brighter than the 3D secondary magnitude limit results in a poorer environment recovery. This means that a deeper sample is favored when constructing the 2D density field.

\section{The Correlation between Environment and Galaxy Property}
\label{sec:5}

In Section \ref{sec:4} we optimized the choice of $N_{\rm 2D}$ for the 2D projected density measurement to yield the best correlation with the 3D real-space environments by using the $r_{s}$ metric. In this section, we use these optimized results to study how the color--density relation in the simulation changes when varying the photo-z uncertainties and outlier rates. Although the density--color and/or halo mass--color relations seen in the simulations may not fully represent the observed Universe, this provides us with a guideline to understand how reliably we can study the dependence of galaxy properties on environment using the photo-z samples.

\subsection{Environment vs. Galaxy Color}

\label{sec:5.1}

To explore the relation between galaxy color and environment, we compare the apparent magnitude $i$ versus $g-i$ colors of galaxies located in the 20\% most dense and the 20\% least dense galaxy environments respectively. We note that although conventionally the color--magnitude relation is defined in the rest frame quantity when studying the color--density relation, here we only look at the observed quantity since the redshift range is very small and our main purpose is to see if the density dependence of color distributions can still be revealed in the photometric redshift sample, rather than quantifying the 'color--density relation' itself. Galaxies are first classified to be red or blue according to their locations in the observed Color--Magnitude Diagram (CMD). We use $g - i = 1.5$ and 1.75 as dividing lines to separate blue and red galaxies at $0.3 < z < 0.5$ and $0.6 < z < 0.8$, respectively. Next we bin the galaxies according to their $i$-band apparent magnitude and then compute the percentage of red galaxies defined as: 

\begin{equation}
	f_{\rm red} = \frac{N_{\rm red}}{N_{\rm bin}},
\end{equation} 
where $N_{\rm bin}$ is the total number of galaxies and $N_{\rm red}$ is the number of red galaxies in each bin. 

We determine the choice of $N_{\rm 2D}$ that yields the largest $r_{s}$ for photo-z samples with different photo-z uncertainties using the methodology described in Section \ref{sec:4.2}. In the case where we study the color--density relation for the environment scale corresponding to the 6$^{\rm th}$ nearest neighbor in the 3D space, i.e., $\rho_{6}$, it is found that the optimized $N_{\rm 2D}$ = 6, 6, 6, 12 is for the cases with photo-z error = 0.00, $0.02(1+z)$, $0.04(1+z)$ and $0.06(1+z)$ respectively (see the bottom panel of Figure \ref{fig:7}). We note that in the case of photo-z error = $0.06(1+z)$, although $N_{\rm 2D}$ = 12 is the best choice for optimization, we still adopt $N_{\rm 2D}$ = 6 to show its CMD for convenience as there is almost no significant difference between using $N_{\rm 2D}$ = 6 or $N_{\rm 2D}$ = 12 for the optimization. 

\begin{figure}
\epsscale{1.0}
\plotone{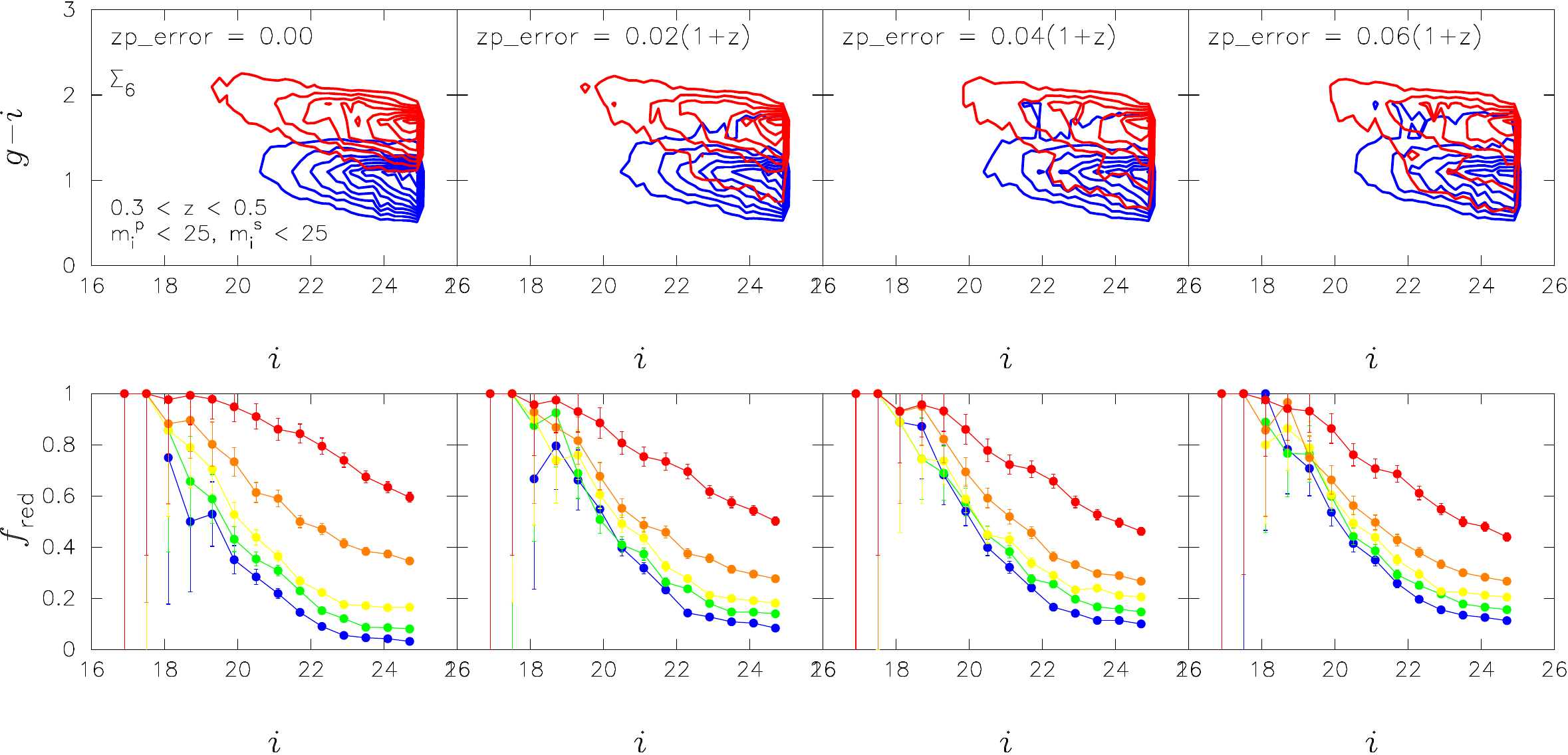}
\caption{Upper panels: color--magnitude diagrams for galaxies in the 20\% most dense (red contour) and 20\% least dense (blue contour) environments with different photo-z errors (from left to right: 0, $0.02(1+z)$, $0.04(1+z)$ and $0.06(1+z)$). Lower panels: the red fraction, $f_{\rm red}$, as a function of $i$-band apparent magnitude with different percentage of density levels: the 20\% most dense (red), 60\% -- 80\% densest (orange), 40\% -- 60\% densest (yellow), 20\% -- 40\% densest (green) and 20\% least dense (blue). The error bars are given by Poisson statistics, and the contours show the regions of constant galaxy number. \label{fig:9} }

\end{figure}

The upper-panels of Figure \ref{fig:9} show the CMD for the 20\% most dense environments (red-contour) and the 20\% least dense environments (blue-contour) for galaxy samples with different photo-z errors. Here we use the 2D projected measurements $\Sigma_{6}$ and set $V_{\rm cut}$ = $\pm0.001(1+z)$, $\pm0.02(1+z)$, $\pm0.04(1+z)$ and $\pm0.06(1+z)$ for the case with photo-z error = 0.00, $0.02(1+z)$, $0.04(1+z)$ and $0.06(1+z)$ respectively. All cases are considered for $m_{i}^{p} < 25$ and $m_{i}^{s} < 25$. Here the contours connect points with equal pixel-density in the CMD. The lower-panels of Figure \ref{fig:9} show the red fraction, $f_{\rm red}$, as a function of $i$-band apparent magnitude for local densities correspond to the 20\% most dense (red), 60\% -- 80\% densest (orange), 40\% -- 60\% densest (yellow), 20\% -- 40\% densest (green) and 20\% least dense (blue). The error bars show the 1-$\sigma$ Poisson uncertainty in each bin. 

It is clear that in the simulation, galaxy colors are strongly correlated with environment: being redder in denser environments. In the case where there is no photo-z error, the red and blue contours occupy distinct regions in the CMD. This trend is in good agreement with observational results \citep{2004ApJ...615L.101B, 2006MNRAS.370..198C, 2007MNRAS.376.1445C, 2007ApJS..172..270C}. As the photo-z error increases (from left to right), red and blue contours begin to overlap. To further quantify the influence of the photo-z uncertainty on the CMD, we plot the red fraction as a function of $i$-band apparent magnitude for galaxies located in 5 different density percentiles (bottom panel of Figure \ref{fig:9}). The difference in the red fraction becomes gradually smaller with increasing photo-z error. Considering the case without photo-z error and galaxies $m_{i} > 20$, the difference in red fraction between the 20\% most dense and 20\% least dense environments ranges from 0.5 to 0.6, but decreases to 0.3 -- 0.4 in the case of photo-z error = $0.06(1+z)$. Nevertheless, it is still encouraging that even in the worst case (photo-z error = $0.06(1+z)$), the dependence of the red fraction on galaxy environment can still be seen.

\begin{figure}
\epsscale{1.0}
\plotone{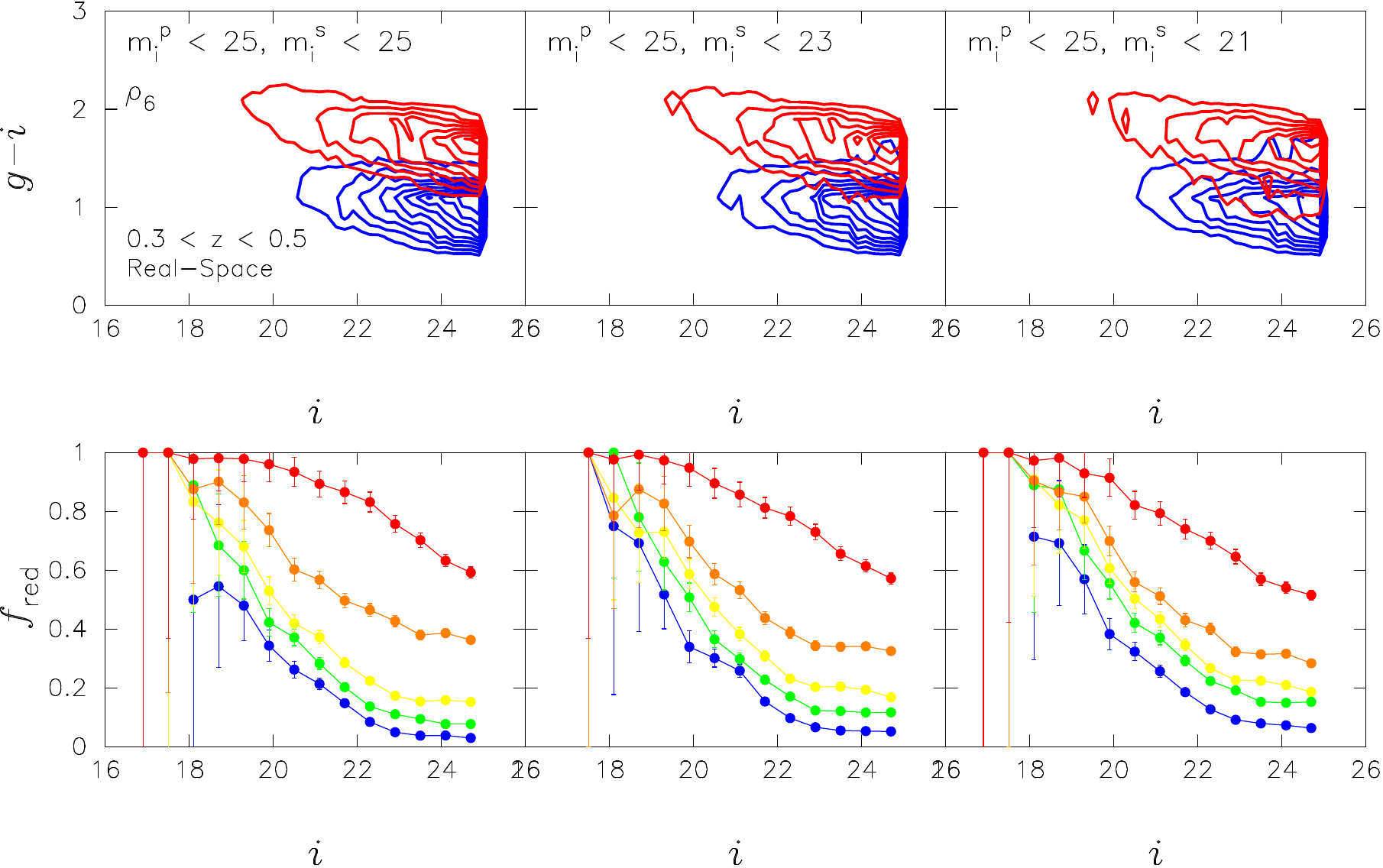}
\caption{Upper panels: color--magnitude diagrams for galaxies in the 20\% most dense (red contour) and 20\% least dense (blue contour) real-space environments with different secondary magnitude limits (from left to right: $m_{i}^{s} < 25$, $m_{i}^{s} < 23$ and $m_{i}^{s} < 21$). Lower panels: the red fraction, $f_{\rm red}$, as a function of $i$-band apparent magnitude with different percentage of density levels: the 20\% most dense (red), 60\% -- 80\% densest (orange), 40\% -- 60\% densest (yellow), 20\% -- 40\% densest (green) and 20\% least dense (blue). The error bars are given by Poisson statistics, and the contours show the regions of constant galaxy number. \label{fig:10} }
\end{figure}

Figure \ref{fig:10} presents similar information to Figure \ref{fig:9}, but now for the cases with various secondary magnitude limits. Here we consider the cases with $m_{i}^{s} < 25$, $m_{i}^{s} < 23$ and $m_{i}^{s} < 21$. All the three cases are considered using $m_{i}^{p} < 25$. To remove additional uncertainties due to projection effects, the density ranking is based on the 3D real-space environments, $\rho_{6}$. Our results show that the environment measured with fainter secondary magnitude limits yield a better correlation with galaxy properties. This result is somewhat expected because the scales of the environment defined by different $m_{i}^{s}$ are different for a given neighbor N, being greater for brighter $m_{i}^{s}$. If the color--density relation is scale dependent, it can lead to the dependence on the adopted $m_{i}^{s}$. The degraded color--density relation with brighter sample in Figure \ref{fig:10} could be due to environmental effects on color being weaker with increasing environment scale. Furthermore, fainter sample is spatially denser and therefore contains more information about environment than a sparse sample does. Including more galaxies in the density estimate thus also help to characterize the environments. We will investigate the correspondence between galaxy environment and dark matter halo mass in the next section.

\subsection{Environments vs. Dark Matter Halo Mass}
\label{sec:5.2}

\begin{figure}
\epsscale{1.0}
\plotone{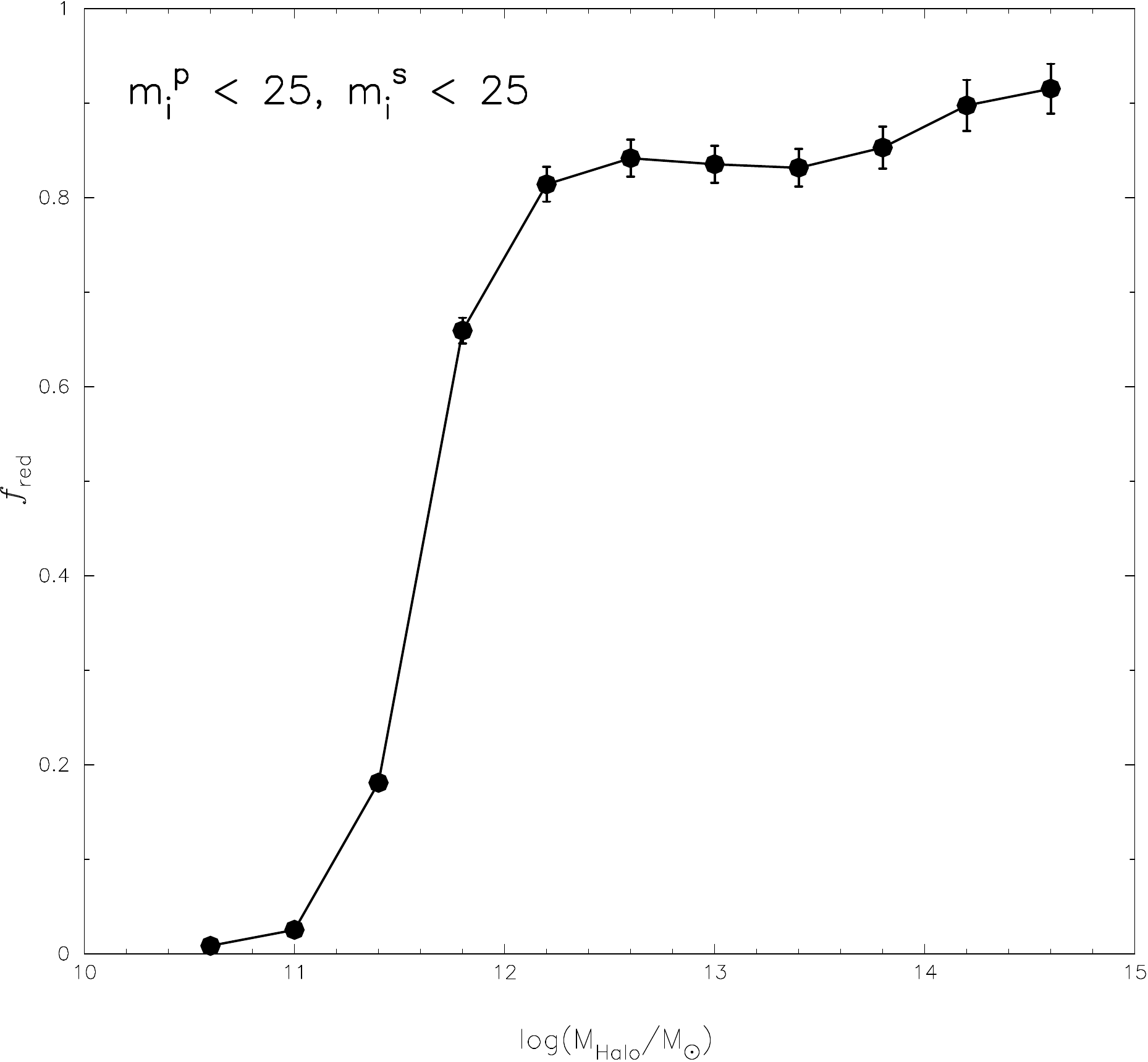}
\caption{Red fraction, $f_{\rm red}$, as a function of dark matter halo mass from galaxy mock catalogs. The error bars are given by Poisson statistics.\label{fig:11} }
\end{figure}

Observationally it is found that galaxies located in massive halos, such as groups and clusters, are in general formed earlier and hence are more evolved, compared to galaxies in the field \citep{2007ApJS..172..284C}. Semi-analytic galaxy formation models have also successfully reproduced the observed trend \citep{1999MNRAS.302..111L, 2000MNRAS.316..107B, 2012MNRAS.419.2133H, 2012MNRAS.419.2670M}. Figure \ref{fig:11} shows the red fraction as a function of dark matter halo mass in our mock catalog based on the model of \cite{2012MNRAS.426.2142L}. As can be seen, the red fraction increases rapidly toward massive halos, in good agreement with observation. 

\begin{figure}
\epsscale{1.0}
\plotone{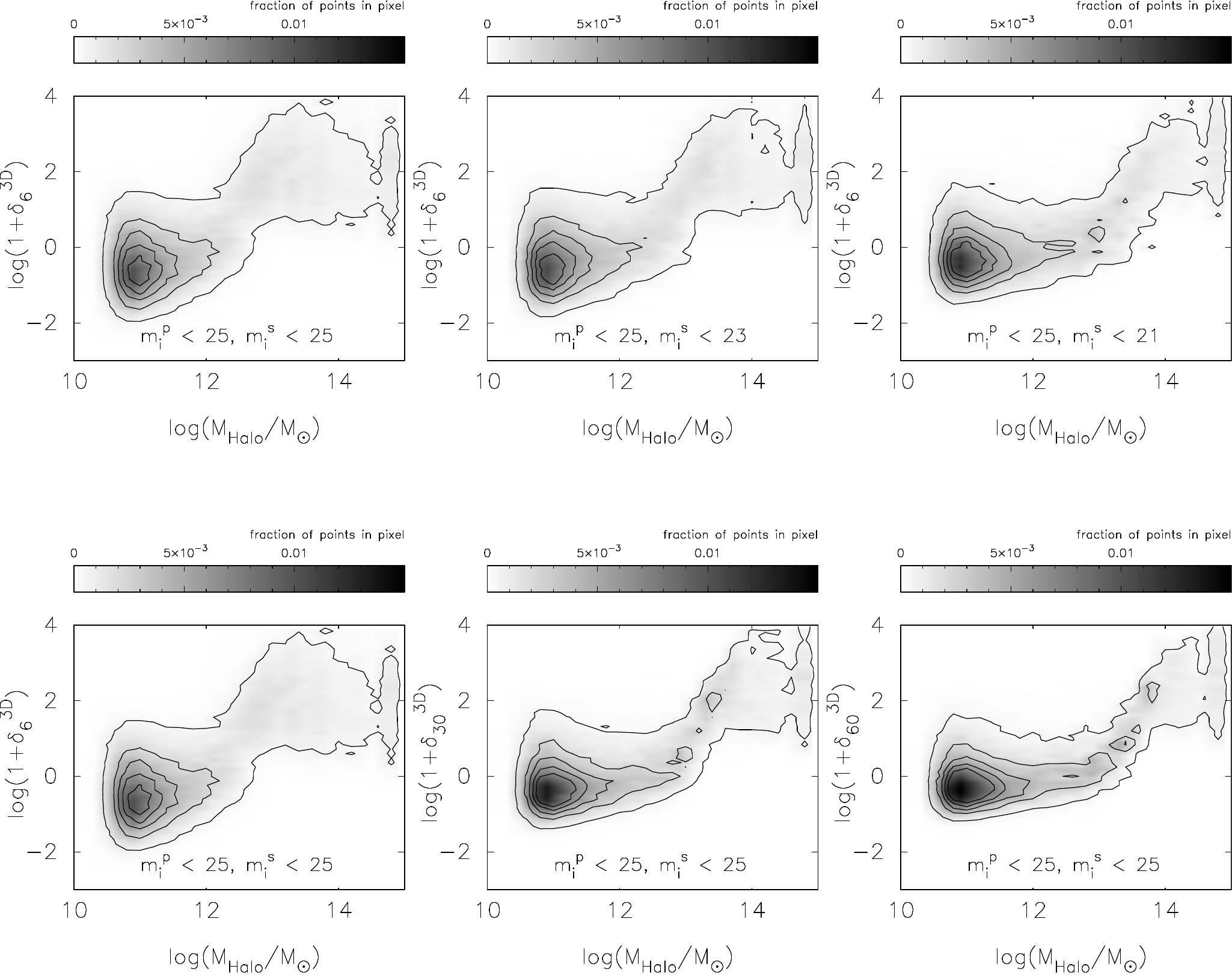}
\caption{Relationship between host halo mass and galaxy overdensity, $1+\delta_{n}^{3D}$ assuming no redshift error, with three different secondary magnitude limits (series of top panels, from left to right: $m_{i}^{s} < 25$, $m_{i}^{s} < 23$ and $m_{i}^{s} < 21$) and choices of $N_{\rm 3D}$ (series of bottom panels, from left to right: $N_{\rm 3D}$ = 6, $N_{\rm 3D}$ = 30 and $N_{\rm 3D}$ = 60). The contours show the regions of constant galaxy number. \label{fig:12} }
\end{figure}

We now proceed to show the correlation between host halo mass and overdensity in order to understand why the galaxy environment calculated using a fainter secondary magnitude limit has a better correlation with galaxy color. We adopt a method similar to the one used in \cite{2012MNRAS.419.2670M} which is to plot the relationship between host halo mass and overdensity. The upper-panels of Figure \ref{fig:12} show the correlations between host halo mass and the 3D overdensity for the galaxy samples with different secondary magnitude limits. In general we find that even in the case of zero redshift error, the 3D environment measures are a poor tracer of mass for individual objects as revealed by the large scatters between 3D density field and halo mass. Similar to what is found by \cite{2012MNRAS.419.2670M}, for low $N_{\rm 3D}$ a galaxy found at high 3D density is actually more likely to be in a low-mass halo than in a high-mass one. Furthermore, our results show that for a fixed $N_{\rm 3D}$, the low density region probed using a brighter $m_{i}^{s}$ has a wider spread in halo mass. In contrast, the low density region measured using a fainter $m_{i}^{s}$ is dominated by galaxies located in small halos ($< 10^{12}$ $\rm M_{\odot}$) where the red fraction drops significantly with decreasing halo mass (see Figure \ref{fig:11}). This is because the scale of the density defined by a fainter magnitude limit for a fixed $N_{\rm 3D}$ is typically smaller and less contaminated by the 2-halo term, and therefore is a better tracer of the halo mass, except for very massive halos ($> 10^{14}$ $\rm M_{\odot}$).

This point is further illustrated in the lower panels of Figure \ref{fig:12} where we show that smaller $N_{\rm 3D}$ yields a stronger correlation between overdensity and host halo mass than using larger $N_{\rm 3D}$. Therefore the tighter relationship between the galaxy colors and densities computed using a fainter secondary magnitude limit seen in Figure \ref{fig:10} can be attributed to the fact that the environment defined using a fainter sample traces more closely with host halo masses.

\begin{figure}
\epsscale{1.0}
\plotone{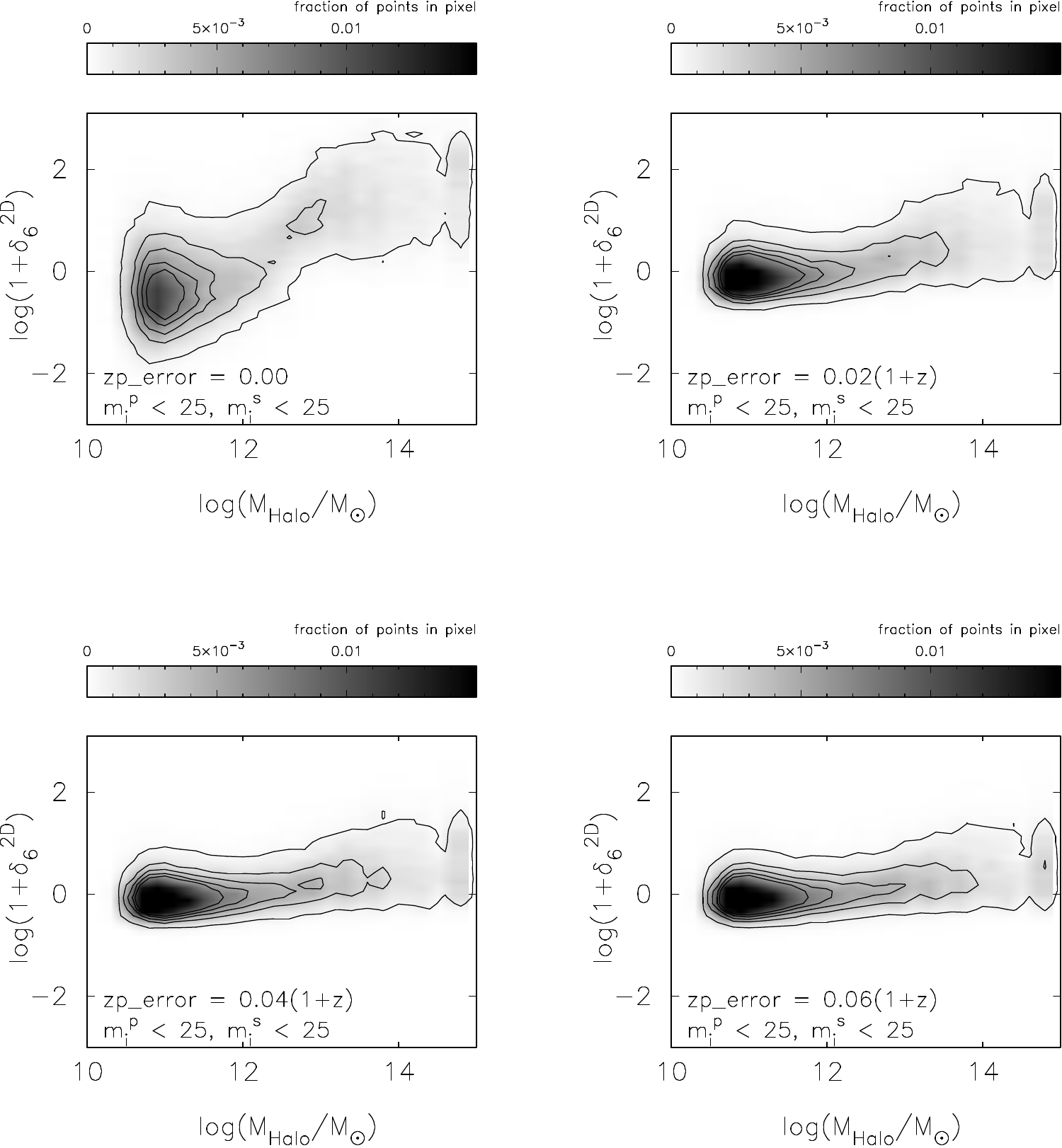}
\caption{Relationship between host halo mass and galaxy overdensity, $1+\delta_{6}^{2D}$, with four different photo-z errors: 0.00, $0.02(1+z)$, $0.04(1+z)$ and $0.06(1+z)$. The contours show the regions of constant galaxy number. \label{fig:13} }
\end{figure}

Figure \ref{fig:13} shows $1+\delta_{6}^{2D}$ versus halo mass for the galaxies with photo-z errors varying from 0.0 to $0.06(1+z)$. The overdensity $1+\delta_{6}^{2D}$ increases with host halo mass but with a large scatter, as seen in \cite{2012MNRAS.419.2670M} which is based on different simulations. Nevertheless, the correlation becomes progressively weaker when photo-z uncertainty increases. In the case of photo-z = $0.06(1+z)$, the correlation is almost flat, suggesting that the 2D projected density is no longer a good tracer of halo mass when photo-z errors are non-negligible.

\subsection{Comparison with Spectroscopic Observation}
\label{sec:5.3}

In previous sections, we present how the photo-z uncertainty can have an impact on the measurement of local density and discuss how well the 2D projected density traces the real-space density when adopting different choices of the size of velocity (redshift) window, magnitude limit, and the $N^{\rm th}$ nearest neighbor. Furthermore, we have also studied how the color--density relation is affected by the presence of photo-z errors and as a function of the magnitude limit that is applied to the secondary sample. In this section, we further investigate the difference in the local density measurements between the photo-z and spectral-z sample using mock galaxies, by taking into account more realistic situations including the incompleteness of the spectroscopic sample.

As we discuss in Section \ref{sec:4}, the photo-z uncertainty has a strong impact on the correlation between the 2D and 3D densities. Ideally, the density measurement based on spectral-z sample is more reliable. However, the spectroscopic observations of galaxies are time-consuming and hence are normally limited to a small sample size, brighter galaxies, and a lower redshift range. Moreover they often suffer from incompleteness due to the limited observing time as well as the fiber and/or slit collisions. In contrast, the photo-z can be relatively easily obtained down to fainter galaxies and out to higher redshifts with a much larger size of sample, but with the drawback that the redshift resolution is substantially poorer compared to spectral-z. Nevertheless, both spectral-z and photo-z samples have been used on environment studies \citep{2006MNRAS.370..198C, 2007A&amp;A...468...33E, 2007ApJS..172..270C, 2012ApJ...744...88Q}. It is thus interesting to investigate to what extend the local density from photo-z samples can be compared to that of the spectral-z samples. To do so, we randomly choose part of the entire spectral-z samples from the mock catalogs to simulate the spectral-z samples with different percentage of completeness. We then apply our optimized scheme as introduced in Section \ref{sec:4.2} to both incomplete spectral-z samples and photo-z samples to compare their $r_{s}$.

\begin{figure}
\epsscale{1.0}
\plotone{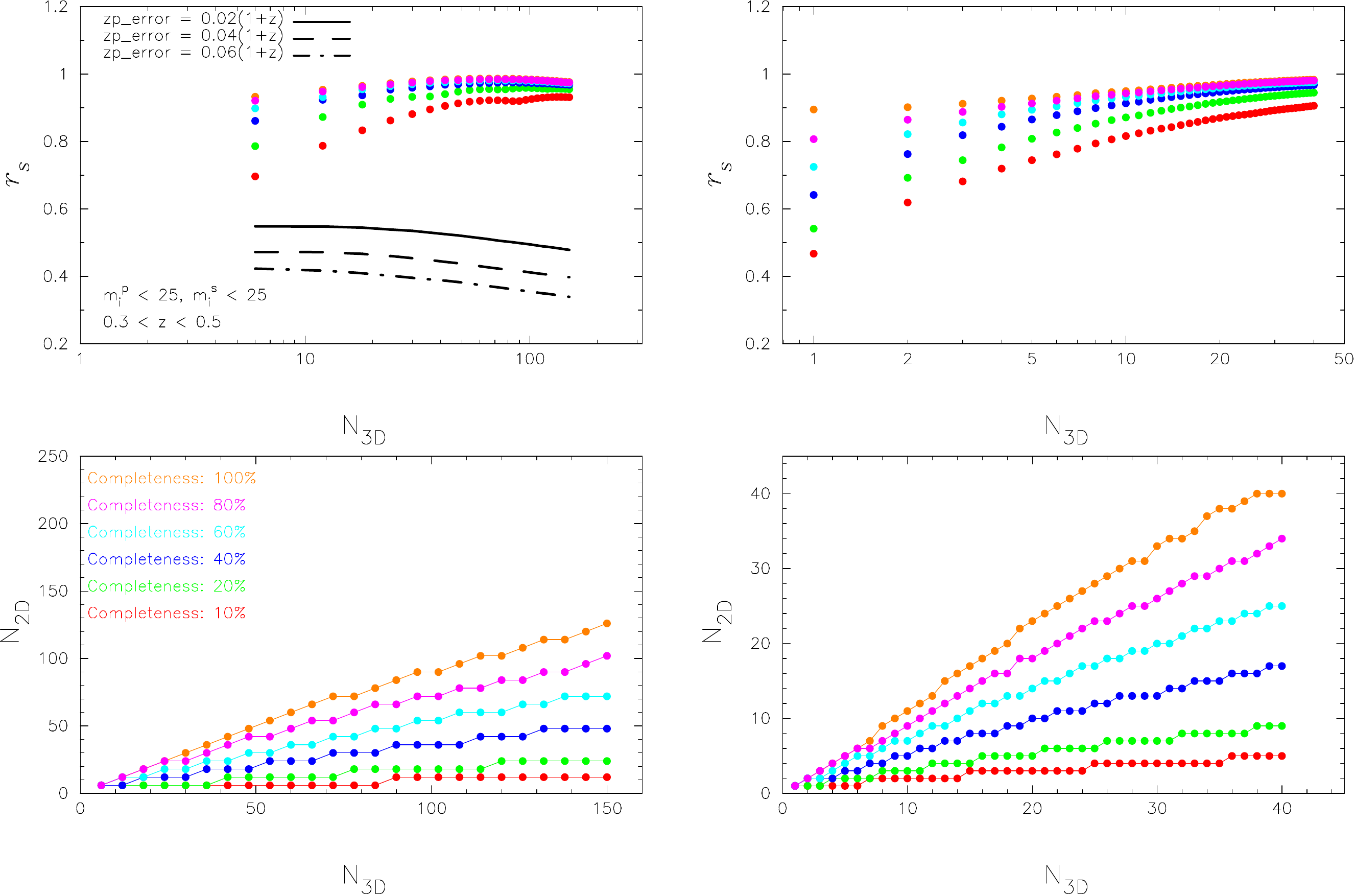}
\caption{The largest $r_{s}$ obtained by varying $N_{\rm 2D}$, for large choice of $N_{\rm 3D}$ (top-left panel) and small choice of $N_{\rm 3D}$ (top-right panel), and the corresponding choice of $N_{\rm 2D}$ (series of bottom panels) that yields the best correlation between the real-space and 2D projected environments for different choices of $N_{\rm 2D}$ as a function of $N_{\rm 3D}$ from mock galaxy catalogs. These cases in which the redshift uncertainty is zero, $m_{i}^{p} < 25.0$ and $m_{i}^{s} < 25.0$ with a wide range of sampling rates. Different colors are for samples with different sampling rate (namely, spectroscopic completeness): 10\%, 20\%, 40\%, 60\%, 80\% and 100\%, and different line-styles represent cases with different photo-z uncertainties as shown in Figure \ref{fig:7}. \label{fig:14} }
\end{figure}

\begin{figure}
\epsscale{1.0}
\plotone{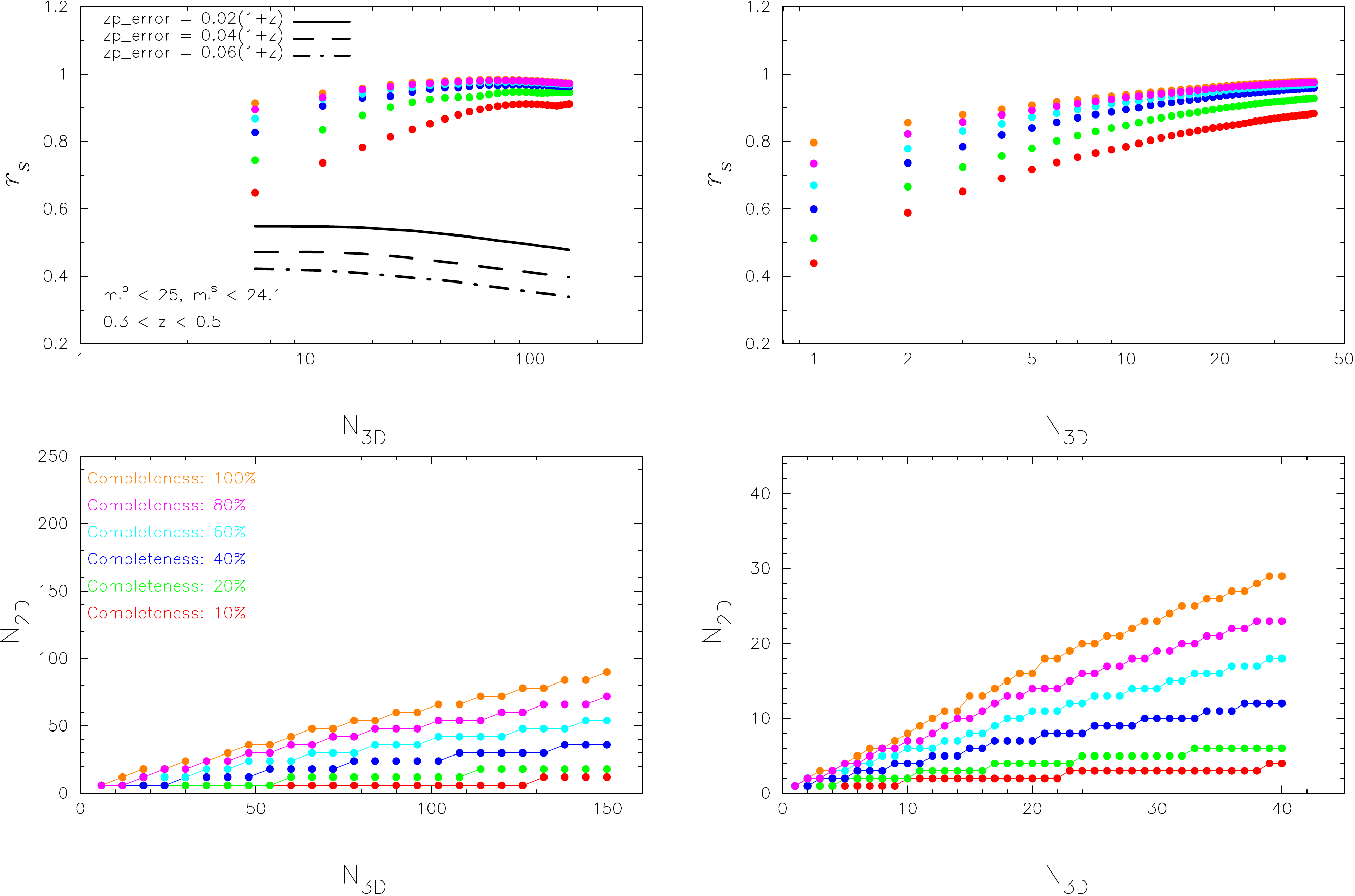}
\caption{Similar to Figure \ref{fig:14} but with $m_{i}^{s} < 24.1$ in the case of the redshift uncertainty equal to zero. \label{fig:15} }
\end{figure}

\begin{figure}
\epsscale{1.0}
\plotone{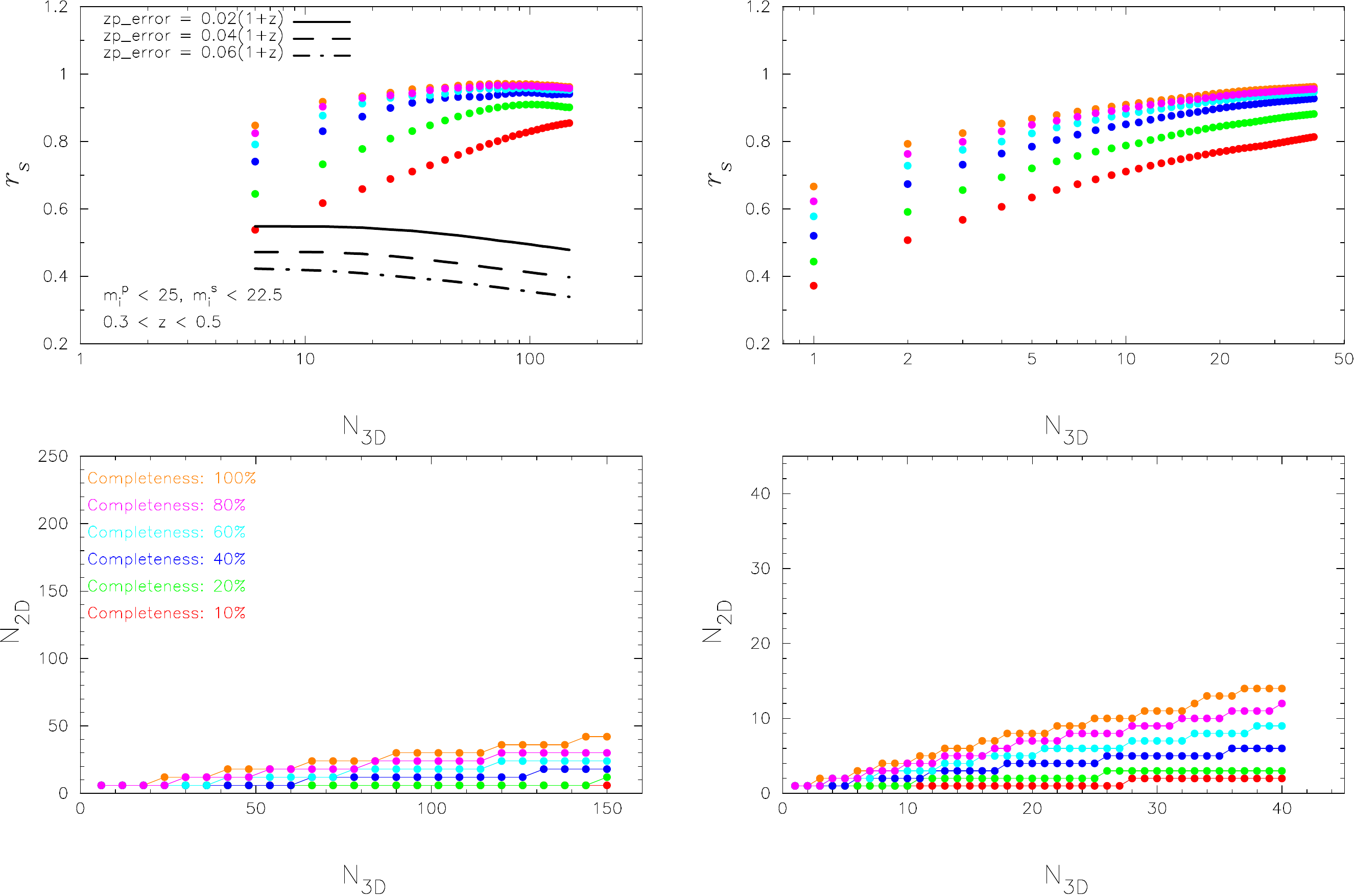}
\caption{Similar to Figure \ref{fig:14} but with $m_{i}^{s} < 22.5$ in the case of the redshift uncertainty equal to zero. \label{fig:16} }
\end{figure}

Figures \ref{fig:14} -- \ref{fig:16} show $r_{s}$ as a function of $N_{\rm 3D}$ for spectral-z samples with different completeness of 10\%, 20\%, 40\%, 60\%, 80\% and 100\% (denoted by different colors). For comparison, we also overplot the results of photo-z sample with different photo-z uncertainties presented in different line styles in the top-left panel. To fairly compare $r_{s}$ on the same physical distance scales among various cases, $r_{s}$ is calculated using the $N_{\rm 3D}$ that corresponds to the same $N^{\rm th}$ nearest neighbor in the case of 100\% complete spectroscopic sample. A zoomed-in version on small scales are shown in the top-right panel. The difference among the three figures (\ref{fig:14} -- \ref{fig:16}) is the secondary magnitude limit applied. For example, in Figure \ref{fig:14} we consider the case where galaxy environments are calculated by using $m_{i}^{s} < 25.0$. It shows that the galaxy environments calculated by using an incompleteness spectral-z sample with this deep magnitude selection are more reliable than those galaxy environments calculated by using a complete photo-z sample. 

However, in general it is difficult to obtain spectroscopic redshifts for a large sample of very faint galaxies. For example, the DEEP2 survey \citep{2013ApJS..208....5N} is limited to $R < 24.1$ and the zCOSMOS bright sample \citep{2007ApJS..172...70L} is limited to $ i = 22.5$. Next we vary the secondary magnitude limits of the spectroscopic sample to see how the trend changes. Here we consider the two optimized results, $m_{i}^{s} < 24.1$ and $m_{i}^{s} < 22.5$, to roughly mimic the results of DEEP2 sky survey and zCOSMOS-bright survey, respectively. Strictly speaking the DEEP2 is limited in the $R$-band instead of $i$-band, however, here we simply adopt the $i$-band in order to reveal the trend more clearly. Our results show that in the case of $m_{i}^{s} < 24.1$, the performance of density recovery with photo-z error as low as $\sim0.02(1+z)$ is always worse than that of the spectral-z samples. However, when the magnitude limit of the spectral-z samples decreases to $m_{i}^{s} < 22.5$, the performance becomes comparable to the spectral-z sample with 10\% completeness. In other words, as the spectral-z sample gets brighter, the $r_{s}$ coefficients between the spectral-z and photo-z samples become closer. However, the difference between their $r_{s}$ coefficients gradually becomes larger when we probe a larger scale of environment, as described in Section \ref{sec:4.2}. That is, a deeper photo-z sample can yield similar performance as good as an incomplete, shallower spectral-z sample, but this is only restricted to small-scale environments.

\subsection{Effect of Outliers}
\label{sec:5.4}

So far the studies on the effect of the photo-z uncertainty on the density measurement are carried out by perturbing the redshifts of mock galaxies with a Gaussian function. However, this method does not totally mimic the realistic case because the photo-z errors may not exactly follow the Gaussian distribution. For example, in the cases where there are not enough numbers and/or wavelength coverage of bandpasses, the feature of the Lyman break ($\sim$ 912 angstrom) can be misidentified as Balmer break ($\sim$ 4000 angstrom) and vice versa, leading to a catastrophic failure in the photo-z estimation, the so-called `redshift outliers' \citep{2013MNRAS.435.2903B}. Next we study how the outlier rate influences the correlation between the 2D and 3D local density measurements.

\begin{figure}
\epsscale{1.0}
\plotone{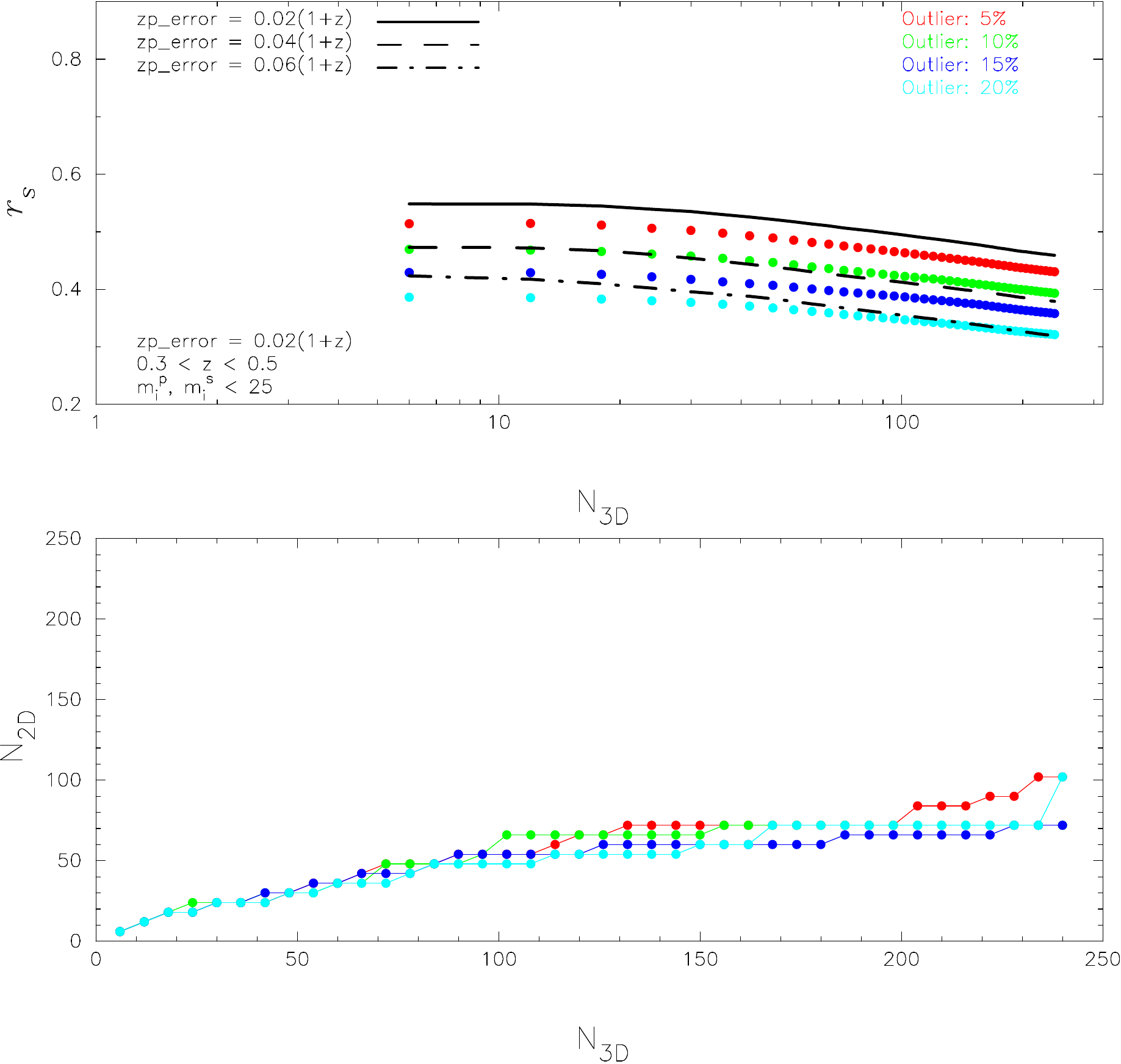}
\caption{The largest $r_{s}$ (upper-panel) obtained by varying $N_{\rm 2D}$ and the corresponding choice of $N_{\rm 2D}$ (lower-panel) that yields the best correlation between the real-space and 2D projected environments for different choices of $N_{\rm 2D}$ as a function of $N_{\rm 3D}$ from mock galaxy catalogs. Different colors represent cases with various outlier rates in the case of photo-z uncertainty equal to $0.02(1+z)$, and different line-styles represent cases with different photo-z uncertainties as shown in Figure \ref{fig:7}. \label{fig:17} }
\end{figure}

To simulate galaxy samples with outliers, we randomly choose part of the entire simulation samples according to the desired outlier rate, and assign them a new redshift randomly between 0 -- 2.0. Figure \ref{fig:17} shows the results using samples with photo-z = $0.02(1+z)$ and four different percentages of outliers: 5\%, 10\%, 15\% and 20\%. Similar to Figures \ref{fig:14} -- \ref{fig:16}, we also mark the results for different photo-z uncertainties in the case of 0\% outliers with different line styles for comparison. From this figure, we can see that $r_{s}$ also strongly depends on the outlier fraction, becoming worse as the outlier fraction increases. The effect is similar to the degradation of photo-z uncertainty and the completeness of the sample. For example, for the sample with photo-z error = $0.02(1+z)$ and outlier rate = 10\% (green dots), we find its optimized result is similar to the result of outlier-free sample with photo-z error = $0.04(1+z)$. 

\begin{figure}
\epsscale{1.0}
\plotone{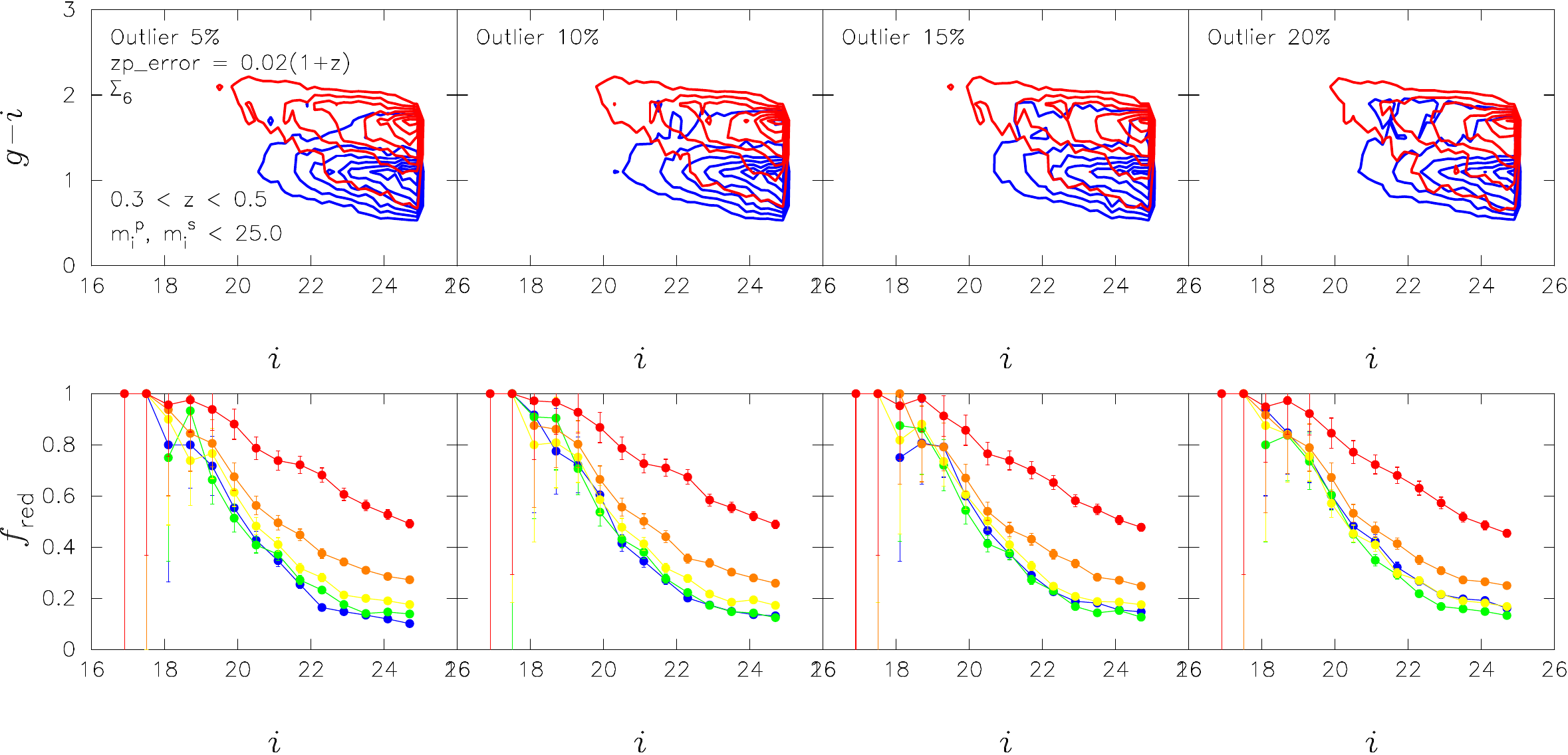}
\caption{Upper panels: color--magnitude diagrams for galaxies in the 20\% most dense (red contour) and 20\% least dense (blue contour) environments with different percentage of outlier rate (from left to right: 5\%, 10\%, 15\% and 20\%). Lower panels: the red fraction, $f_{\rm red}$, as a function of $i$-band apparent magnitude with different percentage of density levels: the 20\% most dense (red), 60\% -- 80\% densest (orange), 40\% -- 60\% densest (yellow), 20\% -- 40\% densest (green) and 20\% least dense (blue). The error bars are given by Poisson statistics, and the contours show the regions of constant galaxy number. \label{fig:18} }
\end{figure}

Figure \ref{fig:18} shows the CMD for the samples with photo-z error = $0.02(1+z)$ and four different percentages of outliers. The environments are measured by using $\Sigma_{6}$, $m_{i}^{p} < 25.0$ and $m_{i}^{s} < 25.0$. As it can be seen, in the case with 5\% outlier, it is comparable to a non-outlier case with photo-z error between $\sim0.02(1+z)$ and $\sim0.04(1+z)$ and the color--density relation can still be revealed. However, as the outlier rate goes up to 10\%, the density measurements for under-dense environments (for example, the least 20\% (blue) and 20\% -- 40\% (green) dense environments) are no longer distinguishable, resulting in a weaker color--density relation. This is in contrast to the situations with pure photo-z errors for which the lowest density curves remain distinguishable. The outliers have larger effects in lower density environments because the change in the density measurements are proportionally larger in those regions when some fraction of galaxies are scatted inside or out the relevant redshift window.

\section{Color--Density Relation for Pan-STARRS1 Data}
\label{sec:6}

The main purpose of this work is to understand the systematics in the 2D density measurement and its limitation, with the ultimate goal of its application to the ongoing and future large photometric surveys. So far we have explored various aspects of the density measurements by using mock galaxy catalogs for which the real-space density is known. We have considered several factors such as photo-z uncertainty, magnitude limit, completeness and outlier rate that make simulation data as similar to realistic samples as possible. However these factors are still not sufficient to imitate the realistic samples. An alternative is to compare the results of the overlapping samples directly between photo-z and spectral-z surveys. We adopt this approach by using the PS1 MD07 photometric redshift catalog \citep{2014ApJ...782...33L} as it covers the well-known EGS field, which has the spectroscopic redshifts from DEEP2, allowing for a direct comparison of the density measurement.

\begin{figure}
\epsscale{1.0}
\plotone{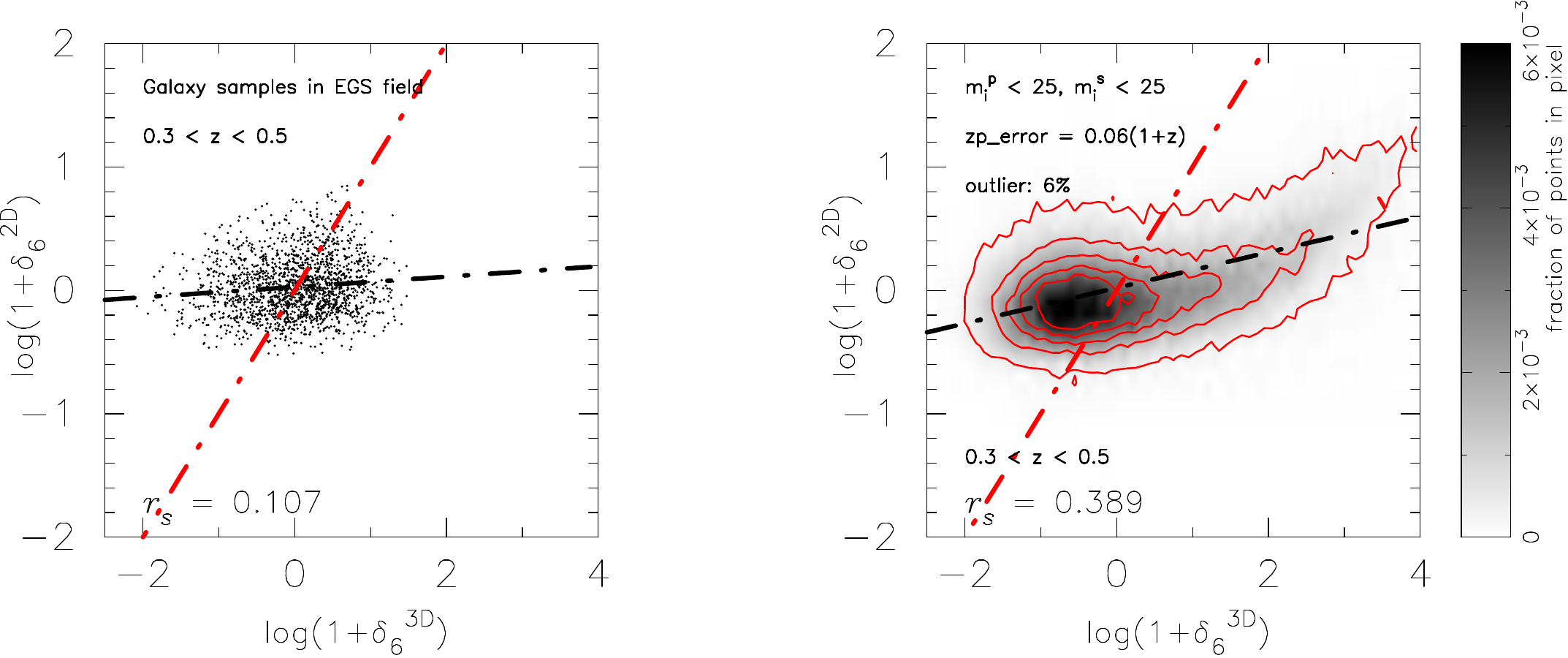}
\caption{Left-panel: The scatter plot of the 3D real-space overdensity, $1+\delta_{6}^{3D}$ versus 2D projected overdensity, $1+\delta_{6}^{2D}$, with galaxy samples in the EGS field. The 2D and 3D environments are calculated by using redshifts from Pan-STARRS1 and DEEP2 respectively. Right-panel: similar to the left panel but using mock galaxy catalogs for the case with real-space overdensity and case with photo-z error = $0.06(1+z)$ and 6\% outliers. The primary and secondary magnitude limits are considered by using $m_{i}^{p} < 25$ and $m_{i}^{s} < 25$. The numbers printed in the bottom-left of each panel indicate the $r_{s}$ coefficient. The black dash-dot lines represent the best fit to the data points, the red dash-dot lines represent the one-to-one relation and the contours show the regions of constant galaxy number. \label{fig:24} }
\end{figure}

We first compute the environments using galaxies with spectroscopic redshifts and compare these environments with those calculated by using photometric redshifts from Pan-STARRS1 for the same galaxies. For comparison, we also utilize the mock galaxy catalogs described in Section \ref{sec:2.1} and perturb their redshifts to simulate the photo-z conditions of Pan-STARRS1 where the typical error is $\sim0.06(1+z)$ and the outlier rate is $\sim 6\%$. Figure \ref{fig:24} shows the scatter plot for galaxies in the EGS field (left-panel) and in the simulations (right-panel). In the left panel, The 2D and 3D environments are evaluated by using samples from Pan-STARRS1 (photo-z) and DEEP2 (spectral-z) respectively. As it can be seen, while we compare the 2D projected and 3D environments in the realistic case, the slope of scatter plot is similar to simulation result but $r_{s}$ is smaller than the results of simulation. While this might be explained by the intrinsic difference in the galaxy spatial distribution between the real Universe and simulations, it is also noticed that the galaxies in the MD07/EGS field span a narrower range in the 3D overdensity because of the smaller field size such that the extreme environments are not well-sampled. As a result, the simulated sample includes very dense environments which are more discernible and easier to be recovered compared to intermediate environments. Nevertheless, there still exists a weak correlation in comparison with real data even though their $r_{s}$ coefficients are smaller than those of the simulated data.

\begin{figure}
\epsscale{1.0}
\plotone{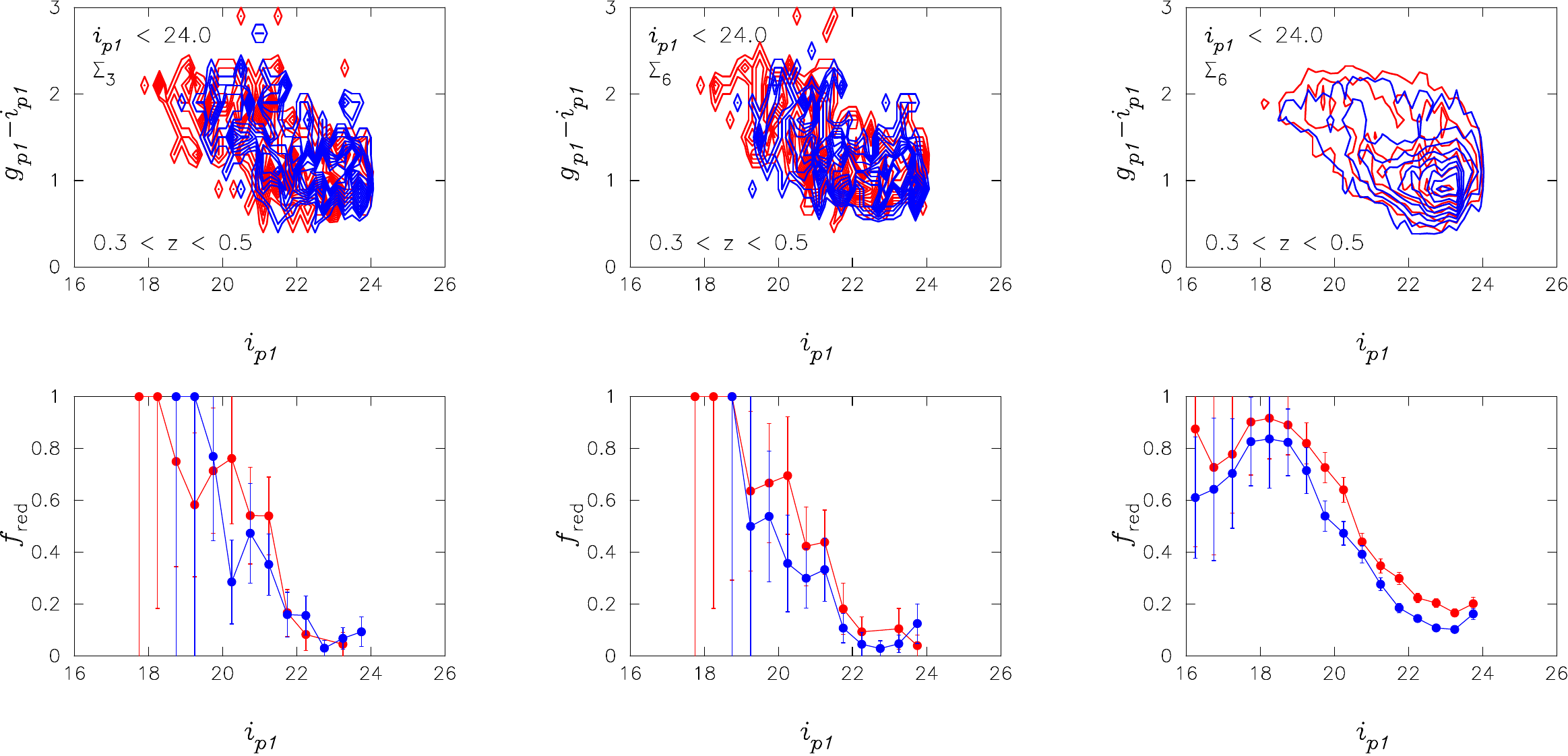}
\caption{Upper panels: the color--magnitude diagrams for galaxies in the 20\% most dense (red contour) and 20\% least dense (blue contour) galaxy environments with different datasets over the redshift interval $0.3 < z < 0.5$ (Left: spectral-z sample from DEEP2 in the EGS field; Mid: PS1 photo-z sample in the overlapping region with the EGS field; Right: PS1 photo-z sample with $\sim5$ $\rm deg^{2}$ ). Lower panels: the red fraction, $f_{\rm red}$, as a function of $i$-band apparent magnitude in the 20\% most dense (red) and 20\% least dense (blue) galaxy environments. The error bars are given by Poisson statistics, and the contours show the regions of constant galaxy number. \label{fig:19} }
\end{figure}

To know whether the color--density relation can still be revealed in the realistic photo-z sample, in the Figure \ref{fig:19} we plot the CMD (upper panels) and the red fractions versus $i_{p1}$ magnitude (lower panels) for galaxies located in the 20\% most dense (red contour) and 20\% least dense (blue contour) galaxy environments. The red fraction is defined as the ratio of the number of galaxies with $g_{p1} - i_{p1}$ color redder than 1.5 to that of the full sample. For each galaxy in the right-panel of PS1, we compute $\Sigma_{6}$ with $m_{i}^{s} < 24$, using the photometric redshifts derived in \cite{2014ApJ...782...33L} with the redshift range of $0.3 < z < 0.5$. The left panel of Figure \ref{fig:19}, which is for comparison, shows the DEEP2 result using galaxy densities computed by \cite{2005ApJ...634..833C}. Their density measurements have been corrected for several effects such as the survey edges, redshift precision, redshift-space distortion and target selection as described in \cite{2005ApJ...634..833C}. Therefore, their measurements can be regarded as the 'true' answer in this comparison. The middle panel shows the result calculated with PS1 photometric redshifts only for galaxies located in the region overlapping with the EGS field which is $\sim0.5$ $\rm deg^{2}$ ($\sim1,500$ galaxies), while the right panel shows the result based on the entire PS1/MD07 field of $\sim5$ $\rm deg^{2}$ ($\sim25,000$ galaxies). To minimize the impact of the edge effects, we also exclude galaxies near the survey boundaries when showing the color--density relation. Among all the three samples, the color--density relation is only significantly detected ($>3\sigma$) in the $\sim5$ $\rm deg^{2}$ PS1 photo-z sample. This is because although the photo-z uncertainty in general contaminates the density measurement, which leads to some systematics in the color--density relation, the random errors can be largely improved given the large volumes probed by a photometric survey. In other words, the reduced errors due to the larger sample are sensitive enough to allow for the detection of a 'degraded' relation between red fraction and environment. 

\begin{figure}
\epsscale{1.0}
\plotone{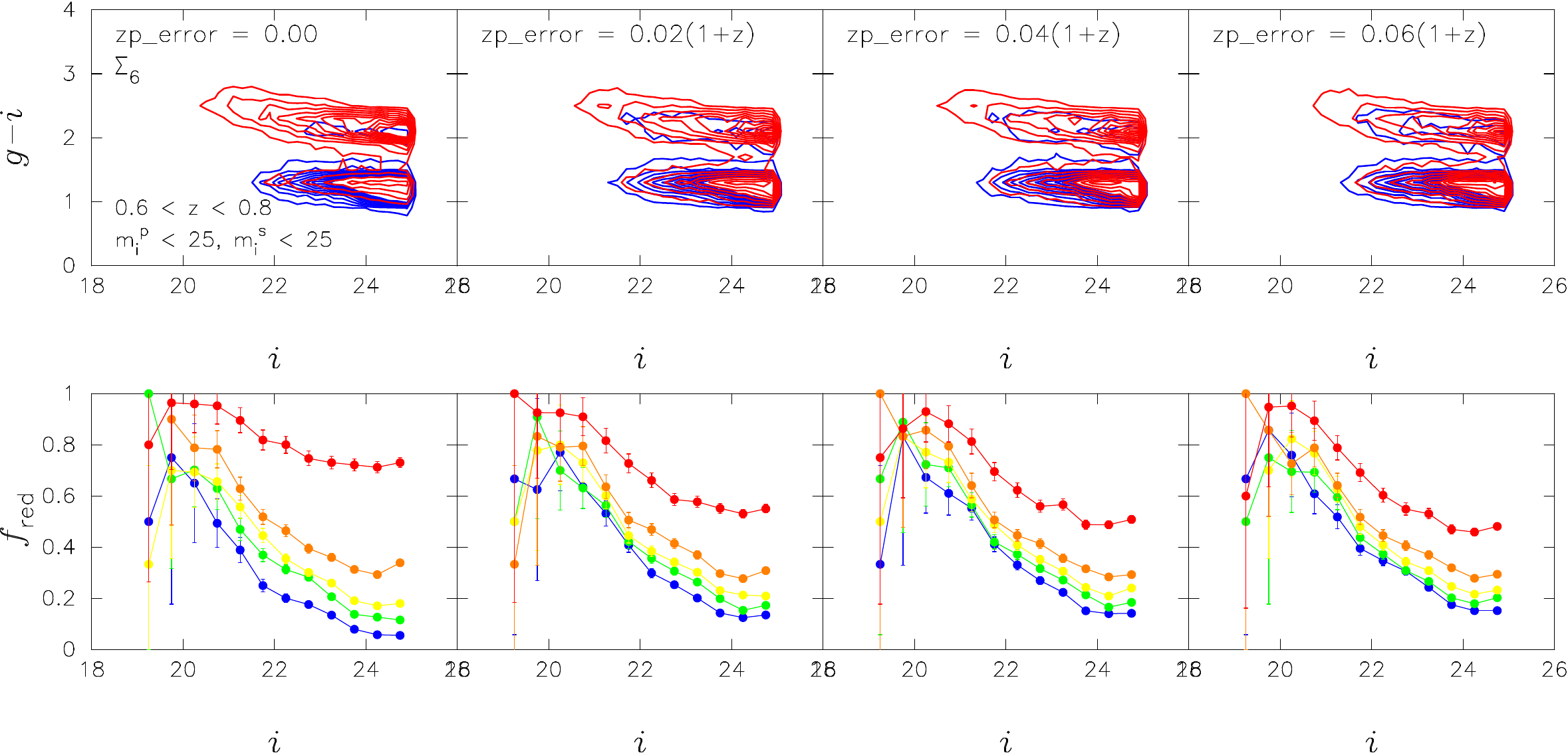}
\caption{Similar to Figure \ref{fig:9}, but for the redshift range of $0.6 < z < 0.8$. \label{fig:28} }
\end{figure}

\begin{figure}
\epsscale{1.0}
\plotone{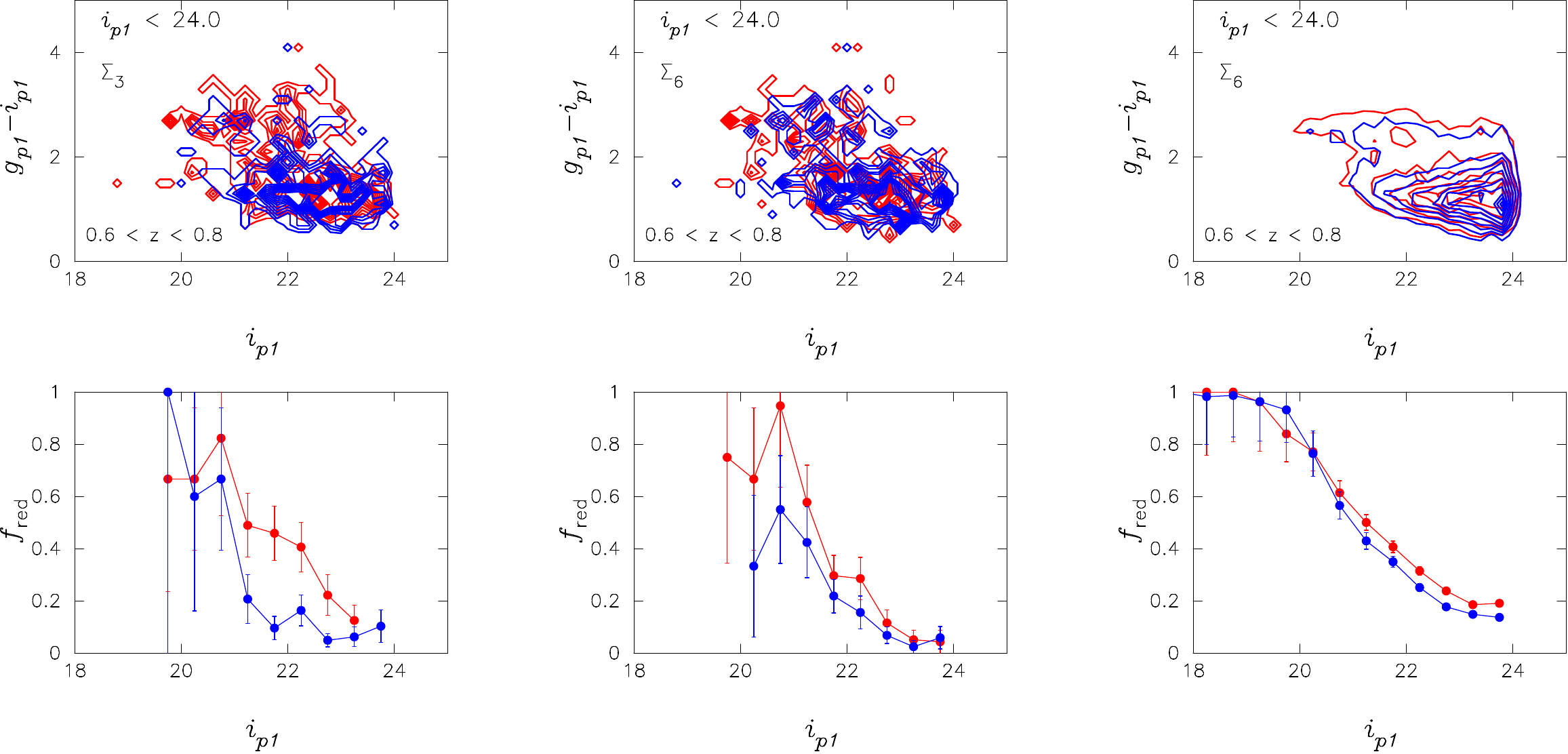}
\caption{Similar to Figure \ref{fig:19}, but for the redshift range of $0.6 < z < 0.8$. \label{fig:20} }
\end{figure}

Furthermore, we also extend our study from low redshift range ($0.3 < z < 0.5$) to higher redshift range ($0.6 < z < 0.8$).  In this redshift bin, $g_{p1} - i_{p1} = 2.0$ is used to separate blue and red galaxies. Similarly we first show the color--density relation at redshift range $0.6 < z < 0.8$ using our simulated dataset (Figure \ref{fig:28}) and PS1 samples (Figure \ref{fig:20}, $\sim45,000$ galaxies). As shown in the right panel of Figure \ref{fig:20}, the difference in the red fraction between two extremely environments is still detectable at $\sim2-3\sigma$ level. The Kolmogorov-Smirnov test \citep{ANDRADE2001909} on the color distributions for galaxies with $21 < i_{p1} < 23$ located in the most 20\% and least 20\% density percentiles returns a value of $p << 0.1\%$, rejecting the null hypothesis that the color distributions of galaxies are drawn from the same population. 

\section{Conclusions}
\label{sec:7}

In this work, we have studied how the 2D projected environment correlates with the 3D real-space environment. Using the Durham mock galaxy catalogs, we investigate various parameters in measuring the 2D projected environment and find the best parameters which maximize the Spearman's rank correlation coefficient, defined as $r_{s} = 1 - \frac{6 \sum d^{2}_{i}}{s(s^{2}-1)}$, which is a quantity for quantifying the correlation between 3D real-space and 2D projected overdensities. When applying the $N^{\rm th}$ nearest neighbor method to the PS1 photo-z sample, we show that color--density relation can still be revealed despite of the sizable photo-z errors inherent in the data. Our main conclusions are as follows.

(i) The correlation between the 2D projected and 3D real-space overdensity is sensitive to the photo-z uncertainty. Smaller $N_{\rm 3D}$ is recommended for photo-z samples to achieve a better correlation (larger $r_{s}$) between 2D projected and 3D real-space environments.

(ii) As the scale of 3D real-space environment increases, the $r_{s}$ derived by using spectral-z and photo-z samples show the opposite trend: the correlation becomes gradually stronger for the spectral-samples but worse for the photo-z samples.

(iii) The 2D projected environment measurements are less sensitive to the redshift interval ($V_{\rm cut}$). The redshift interval comparable to the photo-z uncertainty yields 2D projected overdensity that reasonably traces the real-space density.

(iv) The magnitude limit should also be considered when computing local densities of galaxies. The 2D environments measured with fainter magnitude limits yield better correlation with the 3D real-space environments derived from the same limiting magnitude sample for a fixed $N_{\rm 3D}$. In addition, the color--density relation is more prominent if the density is measured using fainter magnitude limits for a fixed $N_{\rm 2D}$. This is because the overdensity computed with fainter magnitudes proves smaller scales with the same $N_{\rm 2D}$, and traces the hosting halo mass of galaxies better.

(v) Considering the case calculated by using galaxy samples at $0.3 < z < 0.5$ with $m_{i}^{p} < 25.0$ and $m_{i}^{s} < 25.0$, the recovering performance of small-scale environments for photometric redshift samples with redshift uncertainty of $0.02(1+z)$ is roughly comparable to that for shallower $i \sim22.5$ spectroscopic redshift samples with $\sim$ 10\% completeness. In addition, the effect of catastrophic failures in the photo-z measurements on the density measurement is similar to that of the photo-z errors.

(vi) Using Durham mock galaxy catalogs in the redshift range of $0.3 < z < 0.5$, we show that the density--dependent red fraction can still be revealed in photometric redshift samples with photo-z uncertainty up to $0.06(1+z)$. Similarly with photo-z sample from PS1, we show that the color--density relation is also present in the sample whose photo-z uncertainty is $\sim0.06(1+z)$ and the outlier rate is $\sim$ 6\%, but the significance strongly depends on the sample size. Based on the results of PS1 in the two redshift bins ($0.3 < z < 0.5$ and $0.6 < z < 0.8$), we recommend that the survey size should at least exceed $\sim$ 5 $\rm deg^{2}$ in order to yield $>3\sigma$  results. Larger fields will be required in order to reduce the Poisson errors if going to higher-redshifts as the color--density relation is less prominent and the number density of galaxies is reduced.\\

We thank the anonymous referee for very constructive comments, which greatly improve this manuscript. This work is supported by the Ministry of Science and Technology of Taiwan under the grant NSC101-2112-M-001-011-MY2,  MOST103-2112-M-001-031-MY3 and MOST102-2112-M-001-001- MY3. H.-Y. Jian acknowledges the support of NSC101-2811-M-002-075. W.-P. Chen acknowledges the support of NSC102-2119-M-008-001. We thank Michael C. Cooper for providing the DEEP2 environment measurements for our work. We also thank the PS1 Builders and PS1 operations staff for construction and operation of the PS1 system and access to the data products provided. The Pan-STARRS1 Surveys (PS1) have been made possible through contributions of the Institute for Astronomy, the University of Hawaii, the Pan-STARRS Project Office, the Max-Planck Society and its participating institutes, the Max Planck Institute for Astronomy, Heidelberg and the Max Planck Institute for Extraterrestrial Physics, Garching, The Johns Hopkins University, Durham University, the University of Edinburgh, Queen's University Belfast, the Harvard-Smithsonian Center for Astrophysics, the Las Cumbres Observatory Global Telescope Network Incorporated, the National Central University of Taiwan, the Space Telescope Science Institute, the National Aeronautics and Space Administration under Grant No. NNX08AR22G issued through the Planetary Science Division of the NASA Science Mission Directorate, the National Science Foundation under Grant No. AST-1238877, the University of Maryland, and Eotvos Lorand University (ELTE).

\appendix

\section{$N_{\rm 3D}$ vs. $N_{\rm 2D}$ in Environment Measurements}
\label{sec:8.1}

\begin{figure}
\epsscale{1.0}
\plotone{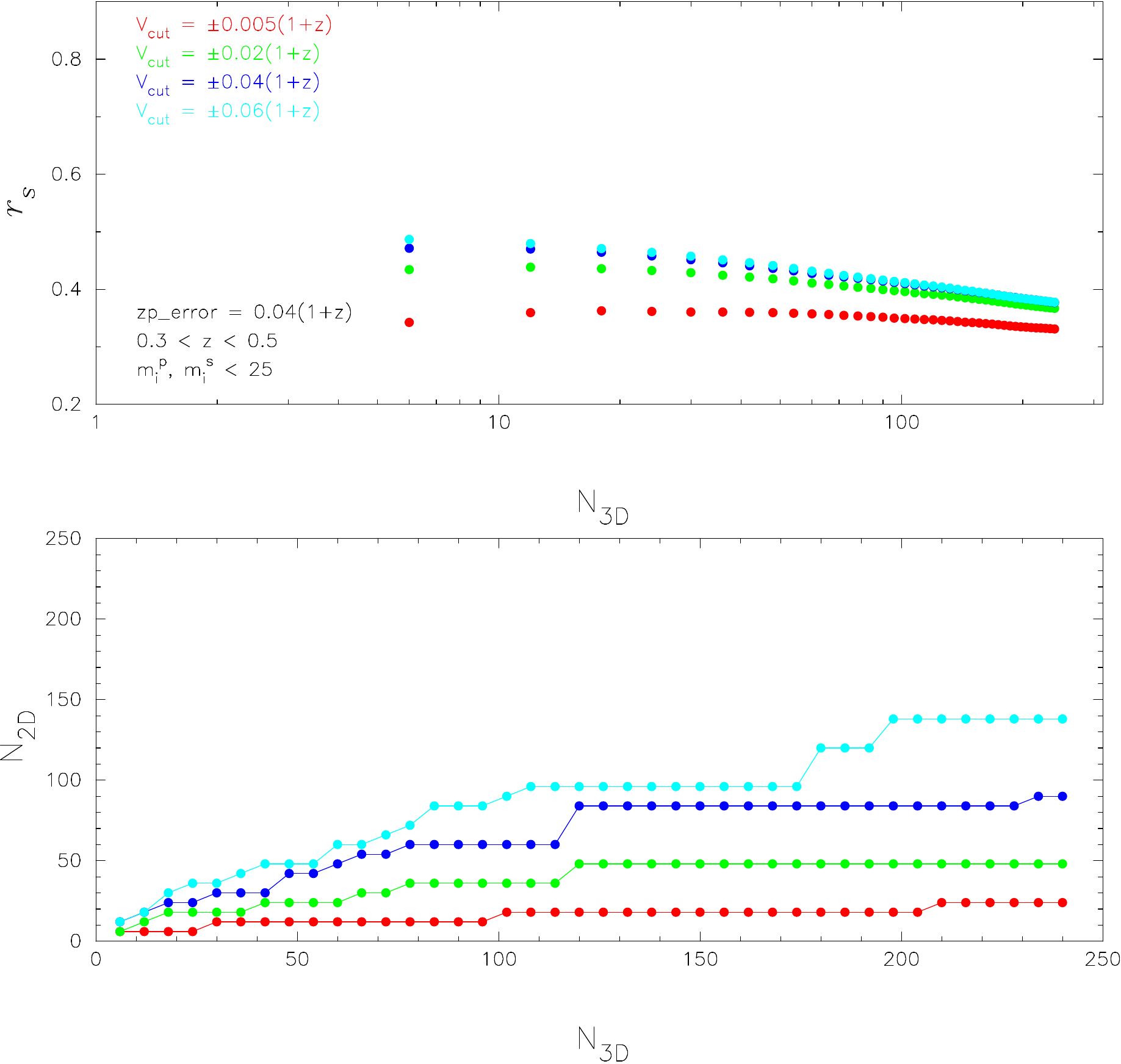}
\caption{The largest $r_{s}$ (upper-panel) obtained by varying $N_{\rm 2D}$ and the corresponding choice of $N_{\rm 2D}$ (lower-panel) that yields the best correlation between the real-space and 2D projected environments for different choices of $N_{\rm 2D}$ as a function of $N_{\rm 3D}$ from mock galaxy catalogs. Different color-dots correspond to samples with different $V_{\rm cut}$ = $\pm0.005(1+z)$, $\pm0.02(1+z)$, $\pm0.04(1+z)$ and $\pm0.06(1+z)$. \label{fig:26} }
\end{figure}

\begin{figure}
\epsscale{1.0}
\plotone{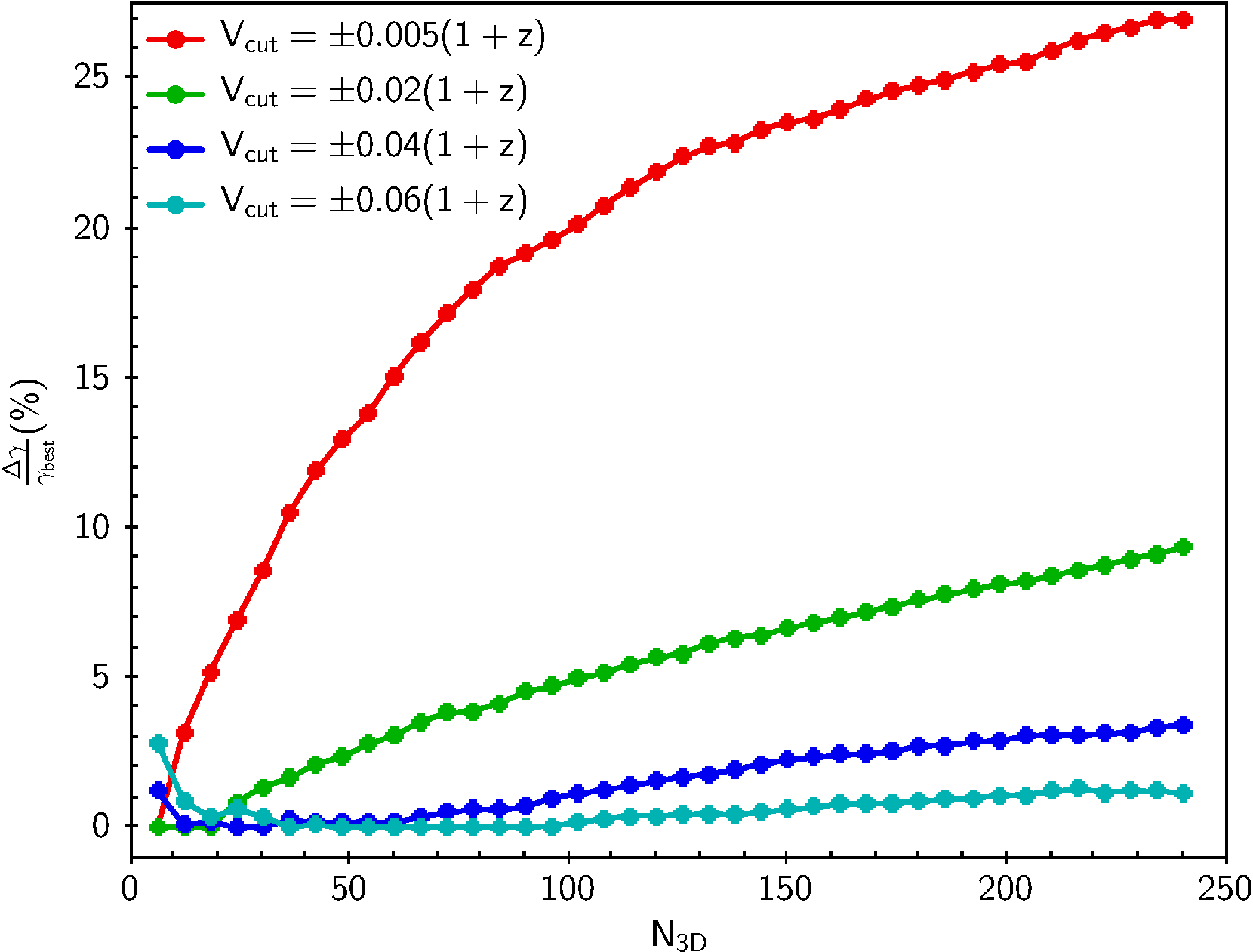}
\caption{The $r_{s}$ difference between the two cases, optimized $N_{\rm 2D}$ and $N_{\rm 2D}$ = $N_{\rm 3D}$, normalized by the former. Red, green, blue and cyan colors are for $V_{\rm cut}$ = $\pm0.005(1+z)$, $\pm0.02(1+z)$, $\pm0.04(1+z)$ and $\pm0.06(1+z)$ respectively. \label{fig:27} }
\end{figure}

As mentioned in Section \ref{sec:4.2}, the optimized $N_{\rm 2D}$ is only half the value of $N_{\rm 3D}$. The ratio of the two quantities in fact depends on the size of the redshift interval when computing the 2D density. Figure \ref{fig:26} shows the optimized scheme for the case with different size of $V_{\rm cut}$. We consider the sample with photo-z error = $0.04(1+z)$ and calculate the environment with different sizes of $V_{\rm cut}$ = $\pm0.005(1+z)$, $\pm0.02(1+z)$, $\pm0.04(1+z)$ and $\pm0.06(1+z)$, corresponding to red, green, blue and cyan dots respectively. As it can be seen, for the cases with size of $V_{\rm cut}$ = $\pm0.02(1+z)$, $\pm0.04(1+z)$ and $\pm0.06(1+z)$, their optimized results are quite similar although the best choices of $N_{\rm 2D}$ are different. The ratio of the optimized $N_{\rm 2D}$ to $N_{\rm 3D}$, roughly, is 1:10, 1:5, 1:2.5 and 1:1.7 for the cases with $V_{\rm cut}$ = $\pm0.005(1+z)$, $\pm0.02(1+z)$, $\pm0.04(1+z)$ and $\pm0.06(1+z)$, respectively. The $N_{\rm 2D}$ in the case of $V_{\rm cut}$ = $\pm0.06(1+z)$ is tripled compared to $N_{\rm 2D}$ in the case of $V_{\rm cut}$ = $\pm0.02(1+z)$. Next we investigate how strong the effect of $N_{\rm 2D}$ is on $r_{s}$. In Figure \ref{fig:27} we plot the difference between the two $r_{s}$, one is evaluated by using best choice of $N_{\rm 2D}$, and another is evaluated by $N_{\rm 2D}$ = $N_{\rm 3D}$, normalized by the former, as a function of $N_{\rm 3D}$. As can be seen, their maximum difference is only $\sim$ 9\%, 3\% and 3\% in the case with $V_{\rm cut}$ = $\pm0.02(1+z)$, $\pm0.04(1+z)$ and $\pm0.06(1+z)$ respectively. This suggests that if the size of $V_{\rm cut}$ comparable to the photo-z uncertainty, the 2D local density measured with $N_{\rm 2D}$ $\sim$ $N_{\rm 3D}$ could be as good as that derived with the optimized $N_{\rm 2D}$.


\clearpage

\clearpage

\clearpage

\clearpage

\end{document}